\documentclass{egpubl}
\usepackage{egsr2024}

\SpecialIssuePaper         

\CGFStandardLicense

\usepackage[T1]{fontenc}
\usepackage{dfadobe}  

\usepackage{cite}  
\BibtexOrBiblatex
\electronicVersion
\PrintedOrElectronic

\usepackage{egweblnk} 

\title[Learning to Rasterize Differentiably]%
      {Learning to Rasterize Differentiably}

\author[C. Wu, H. Mailee, Z. Montazeri \& T. Ritschel]
{\parbox{\textwidth}
{\centering 
C. Wu$^{1}$\orcid{0009-0006-9291-4484},
H. Mailee$^{2}$\orcid{0000-0001-8762-055X},
Z. Montazeri$^{2}$\orcid{0000-0003-0398-3105} and
T. Ritschel$^{1}$\orcid{0009-0006-4660-7790}
}
        \\
{\parbox{\textwidth}{\centering 
$^1$University College London, United Kingdom\\
$^2$University of Manchester, United Kingdom
}
}
}

\begin{document}

\begin{acronym}
    \acro{AD}{Automatic Differentiation}
    \acro{CDF}{Cumulative Density Function}
    \acro{MC}{Monte Carlo}
    \acro{MLP}{Multi-layer Perceptron}
    \acro{PBR}{Physically-based Rendering}
    \acro{PDF}{Probability Density Function}
    \acro{CV}{Control Variates}
    \acro{MAML}{model-agnostic meta-learning}
    \acro{SGD}{stochastic gradient descent}
\end{acronym}


\maketitle
\begin{abstract}
    Differentiable rasterization changes the standard formulation of primitive rasterization ---by enabling gradient flow from a pixel to its underlying triangles---  using distribution functions in different stages of rendering, creating a ``soft'' version of the original rasterizer.
    However, choosing the optimal softening function that ensures the best performance and convergence to a desired goal requires trial and error.
    Previous work has analyzed and compared several combinations of softening.
    In this work, we take it a step further and, instead of making a combinatorial choice of softening operations, parameterize the continuous space of common softening operations.
    We study meta-learning tunable softness functions over a set of inverse rendering tasks (2D and 3D shape, pose and occlusion) so it generalizes to new and unseen differentiable rendering tasks with optimal softness. \\


\end{abstract}  

\section{Introduction}

\begin{figure}[t]
  \centering
  \hspace{-0.5cm}
  \includegraphics[width=.48\textwidth]{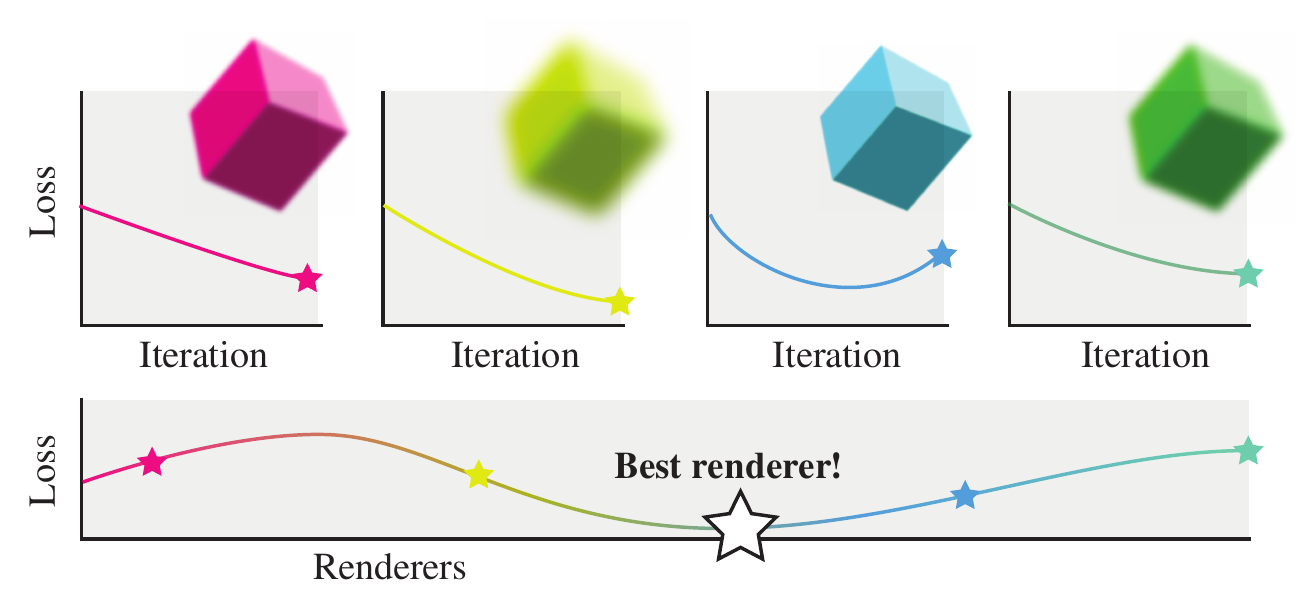}
  \caption{\label{fig:teaser}
  Among the continuously differentiable rasterization renderers, we identify the one most suited to solving a family of inverse rendering tasks.
}
\end{figure}

While forward rendering generates a 2D image based on 3D scene parameters, inverse rendering optimizes these parameters to reproduce a given 2D reference image. The use of modern gradient-based optimizers in this context requires the rendering process to be differentiable, facilitating applications such as model reconstruction \cite{Kato2018mesh}, pose estimation \cite{Loper2014OpenDR, Gupta2020pose}, and the estimation of lighting and materials \cite{Wenzheng2021DIB, Nimier-David2021material}. However, differentiation is challenging due to the discontinuities commonly present in rendering.

The renderer can generally be differentiated in two main ways: either by approximating the gradients using the exact forward rendering process, which may require manually designed gradients or by directly approximating the forward rendering process to enable \ac{AD}. For gradient approximation, Loper et al. \cite{Loper2014OpenDR} leverage the differences between neighboring pixels, while Kato et al. \cite{Kato2018mesh} employ a hand-crafted function. Li et al. \cite{Li2018MC} proposed the integration of gradients using Monte Carlo ray tracing. In contrast, other studies achieve natural differentiability through probabilistic perturbation, such as \cite{Rhodin2015AVS}, which uses Gaussian distributions to blur rasterization and approximates the rasterized value as a density parameter controlling the transparency of blobby objects. Subsequent studies have explored using functions like the square root of a logistic distribution \cite{liu2019soft} and logistic distribution \cite{petersen2021learning} in similar ways. Another approach involves smoothing the cost function via Monte Carlo convolution across optimization parameters \cite{fischer2022plateau}.

These softer rasterization variants have enabled effective inverse rendering by overcoming the challenges posed by discontinuities. However, identifying the appropriate softness function remains a significant challenge. We argue that no single function is universally optimal; the choice of softness depends on the specific problem, and adapting this choice should be automated and systematic.

In this work, we propose a principled approach that addresses this issue at a higher level of abstraction as it is visualized in Figure~\ref{fig:teaser}. Instead of focusing on individual inverse rendering problems, we consider the entire spectrum of potential problems. Utilizing a training set of inverse rendering challenges, we identify the most effective modifications to the renderer in terms of convergence speed and/or quality, making it task-specific and differentiable.

Concretely, we introduce a meta-learning strategy to learn the optimal edge and occlusion softness for a differentiable rasterizer in the form of an \ac{MLP}. We demonstrate that our method surpasses state-of-the-art techniques that all rely on manually selected softness scales for given parametric distributions.

To facilitate further research, we publish the code at \href{https://github.com/Theo-Wu/MetaRas}{https://github.com/Theo-Wu/MetaRas}.

\section{Background}

Differentiable rendering allows computing gradients of 3D scene parameters with respect to the image pixels. 

\subsection{Problem setting}

\mymath{\hardRenderer}{\mathcal R_\mathrm h}
\mymath{\softRenderer}{\mathcal R_\mathrm s}
\mymath{\sceneParameters}{\theta}
\mymath{\optimalSceneParameters}{\sceneParameters^\star}
\mymath{\optimizer}{\mathtt{opt}}
\mymath{\targetImage}{I}
\mymath{\softRendererParameters}{\phi}
\mymath{\optimalSoftRendererParameters}{\softRendererParameters^\star}
\mymath{\stepScene}{m}
\mymath{\stepRenderer}{n}
\mymath{\stepSize}{\lambda}
\mymath{\stepSizeScene}{\stepSize_\sceneParameters}
\mymath{\stepSizeRenderer}{\stepSize_\softRendererParameters}
\mymath{\step}{h}
\mymath{\soft}{s}
\mymath{\distance}d
\mymath{\depth}z
\mymath{\loss}{\mathcal L}
\mymath{\sampleImage}{I_i}

Let \hardRenderer be a common hard renderer that takes scene parameters \sceneParameters and maps them to an image.
Differentiating this function is not possible due to the discontinuity of the parameters and gradients that are typically zero. Hence, the subscript $h$ is used to denote ``hard''.
Formally, let \optimalSceneParameters be the optimal scene parameters for an image \targetImage and let us assume they are unique.
Then, unfortunately,
\smallskip
\[
\optimizer(
||
\hardRenderer(\sceneParameters)-I||_2, 
\sceneParameters)
\neq
\optimalSceneParameters,
\]
where $\optimizer()$ is an optimizer, such as gradient descent that minimizes the first argument (here, the image difference of rendering and reference image) by changing the second argument (here, the scene parameters). 

A differentiable soft renderer \softRenderer, however, would potentially converge to the global optimum, as in: 
\smallskip
\[
\optimizer(
||
\softRenderer(\sceneParameters)-I||_2, 
\sceneParameters)
=
\optimalSceneParameters
\,.
\]\par
Although \softRenderer and \hardRenderer are not identical, the crucial insight is that as long as the gradient-based optimizer converges to the same minimum using its gradients, the specific function used may not significantly impact the outcome. This introduces the possibility of replacing \hardRenderer with a differentiable function, \softRenderer, that results in similar optimal parameters.

In general, \softRenderer and \hardRenderer are not identical; thus, we change the target function we aim to optimize. Nonetheless, the differentiable (soft) version maintains mostly non-zero gradients, fostering optimism that these gradients will converge toward the true, desired optimum. A fundamental insight of this work is the notion that the specific form of \softRenderer and \hardRenderer does not crucially impact the optimization outcome, as long as a gradient-based optimizer can effectively use their gradients to converge to the same minimum.

This leads us to an important question: if we have the flexibility to modify the renderer, how should we systematically determine the best settings? Our goal is to substitute the original function with a differentiable variant that yields similar optimal results when subjected to gradient-based optimization.

The two dominant rendering techniques are path tracing and rasterization. Differentiable Monte Carlo path-tracing is adept at managing all types of illumination given sufficient computational resources, as discussed in prior studies \cite{Li2018MC, Zhang2020pathspace}, without needing explicit boundary sampling \cite{bangaru2020warpedsampling} or sampling silhouette edges \cite{Loubet2019Reparameterizing} by approximating the pre-filtered gradient \cite{Yang2022AAF} to handle discontinuities. Our focus in this paper, however, is on differentiable rasterization. This specific renderer type is restricted to direct illumination but offers the advantage of efficient computation.

\subsection{Differentiable rasterization}

In this section, we first discuss rasterization and then explain its differentiable variant.

\mypara{Hard rasterization}
Rasterization generally refers to the process of determining which pixels fall within a geometric primitive that is projected onto the image plane. This initial step does not involve assigning attributes to each pixel; such tasks are usually handled at the fragment level by dedicated hardware or software. Specifically, rasterization involves processing a set of 2D triangles, each with defined depth and attributes such as color at its vertices, to determine the attributes of each pixel in a 2D image. A common method employs an \emph{edge function} \cite{Pineda1988APA} to test whether a pixel lies within the boundaries defined by the edges of the triangle. If a pixel is determined to be within these boundaries, its attributes are unconditionally assigned based on its position relative to the triangle. This assignment uses a Heaviside step function at the triangle edges, rendering it non-differentiable; the gradient is zero across the field except at the boundaries, where it is undefined due to the abrupt change.

Furthermore, in scenarios where multiple primitives overlap on the same pixel, the attribute selected for that pixel corresponds to the one associated with the closest primitive. This selection process, known as $z$-buffering, ensures that only the attribute of the closest primitive is retained, while the attributes of the other overlapping primitives are discarded. However, similar to other aspects of rasterization, the $z$-buffering process is also non-differentiable.

This indicates that differentiating rasterization involves addressing both occlusion and edge tests. This process essentially reduces to differentiating a function based on:
$\step(\distance)\in\mathbb R\rightarrow\mathbb [0,1]$ 
where \distance represents a \emph{distance} measurement, which could pertain to either 2D spatial dimensions or depth within the scene.

\mypara{Soft rasterization}
The fundamental concept behind soft rasterization, as introduced by Liu et al. \cite{liu2019soft}, involves transforming the traditional hard step function \step into a soft function $\soft(\distance)\in\mathbb R\rightarrow\mathbb [0,1]$ with global support, ensuring non-zero gradients throughout. The output of this soft function is utilized in alpha-compositing, also considering depth attributes. Examples of such functions include Gaussian distributions \cite{Rhodin2015AVS}, the square-root of logistic distributions \cite{liu2019soft}, and exponential functions \cite{chen2019_dibr}.

Petersen et al. \cite{Petersen2019Pix2Vex} proposed a method to soften the $z$-buffer using a weighted softmax function defined over depth values. Additionally, dedicated aggregation functions have been proposed for silhouette computation; these functions differentiate scenes using binary color and are independent of depth. Subsequently, \cite{Petersen2022GenDR} further refined these concepts by defining them as T-conorms and exploring various implementations. Their research demonstrated that several functions could be effective as long as they are monotonous, and it analyzed the performance of each. 

This line of work serves as another important inspiration for our approach, in which we transition from fixed function choices to a task-specific, continuously optimized selection of soft functions. Our approach does not deeply analyze the mathematical properties of these functions; rather, it focuses on fitting the data to practical scenarios, where the primary benefit is reduced computational cost in optimization processes, even without a comprehensive understanding of the underlying reasons.

Furthermore, \cite{liu2019soft} introduced an aggregation function, $A(d,z)$, which softens both spatial distance and depth, allowing gradients to influence both visible and occluded primitives and their $z$ coordinates effectively.

Similarly, Laine et al. \cite{nvidia2020} defined the soft function with local support on surface coverage instead of distance to the edge. They approximate the coverage by the position of edges' crossing points between adjacent pixel pairs. This can be seen as a variant of the truncated linear function on a transformed space. 

To be systematic, we control variables in our comparison and only compare with our backbone GenDR, where everything is consistent except for the MLP to eliminate the influence of other implementations.
Nevertheless, the idea of meta-learned softness is independent of differentiable rasterizer implementation. For example, meta-learning an MLP to replace the linear blending operation in Nvdiffrast \cite{nvidia2020} might also improve it.

For a study of differentiable rasterization in general we refer the readers to the survey by \cite{Kato2020survey}. 

\section{Meta-learning a differentiable rasterizer}

\subsection{Meta problem setting}

To make systematic progress we move the problem to another level of abstraction.
We phrase the challenge as finding a renderer \softRenderer, parameterized in some way by ${\optimalSceneParameters_i}$, that converges best over a set of tasks ${\targetImage_i}$, where $i$ refers to the $i_{th}$ category of task:
\smallskip
\[
\argmin{\softRenderer}
\mathbb E_i
[
||
\optimizer(
||
\softRenderer(\sceneParameters)-I_i||_2, 
\sceneParameters)
-
\optimalSceneParameters_i
||
]
\,.
\]
\smallskip
where \optimizer refers to an optimizer.

In practice, sometimes the ground truth parameters ${\optimalSceneParameters_i}$ are not available; instead, we often have only the images ${\targetImage_i}$. Under these circumstances, the challenge can be reformulated as follows:
\smallskip
\[
\argmin{\softRenderer}
\mathbb E_i
[
||
\softRenderer(
\optimizer(
||
\softRenderer(\sceneParameters)-I_i||_2, 
\sceneParameters))
-
I_i
||
]
\,.
\]
\\
We do not confine ourselves to a predefined, discrete set of functions for manual selection. Instead, we explore the continuous space of all possible soft renderers using meta-learning, which is explained as follows.

\subsection{Meta-learning}

Meta-learning, often described as "learning to learn", is an algorithmic approach that enhances a model by observing how different models perform across various tasks. The \ac{MAML} algorithm was introduced by Finn et al. \cite{finn2017MAML}, which has shown success in learning new tasks with limited training samples by utilizing a double-loop training process.

Given the unique aspects of our tasks, we employ a method similar to \ac{MAML}, as outlined in Algo.~\ref{algo:algo}. Our approach operates with nested loops (L2 and L5): the outer loop adjusts meta-parameters---the shape of the softening functions---while the inner loop, using the Adam optimizer (L8), optimizes the scene settings parameters. Importantly, the inner loop is designed to unfold into a formula that the outer loop can differentiate automatically.
The outer loop's gradient update (L11) modifies the parameters governing the inner optimization, in our case, the differentiable renderer (L12). This ensures that the optimization converges more closely to its target with each iteration.

\begin{figure}[]
  \centering
  \hspace{-0.5cm}
  \includegraphics[width=.48\textwidth]{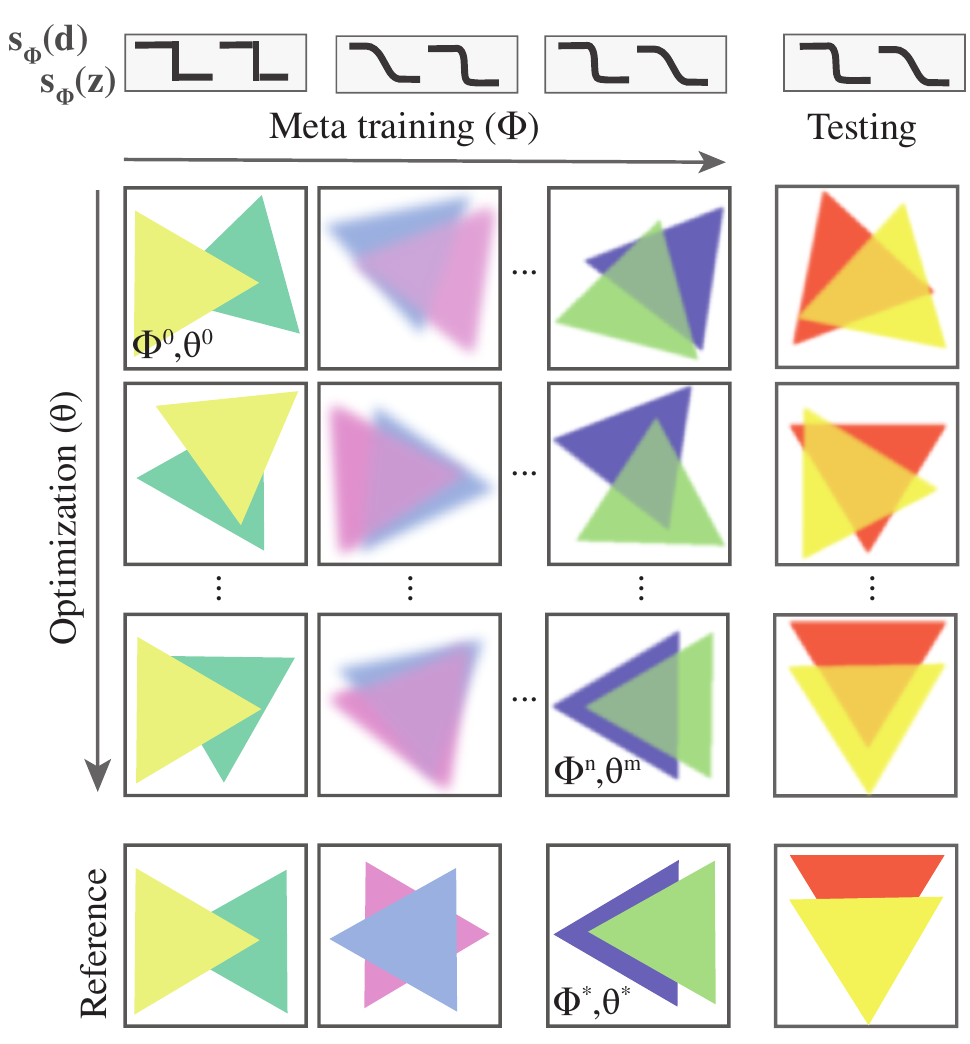}
  \caption{\label{fig:meta-learning}
        \textbf{Meta-learning.} 
        Meta-optimization consists of two training loops to jointly optimize the scene parameters \sceneParameters for one task instance (vertical) and the renderer parameters \softRendererParameters across many task instances (horizontal).
        All columns represent the same task category of changing two triangles' positions to match the reference image in the last row. 
        At test time, the optimal renderer is good at solving unseen tasks, as shown in the rightmost column.
        This is, because, towards the end of meta-training, the optimization itself mimics the reference closely.
        The top shows the soft functions used to render one column: a soft depth step makes the triangles transparent, and a soft edge function makes the edges blurry.
}
\end{figure}

Our implementation diverges from traditional \ac{MAML}, as shown in Figure~\ref{fig:meta-learning}, since we do not apply the learned parameters from our meta-loop to new test instances like classic \ac{MAML}.
Furthermore, rather than accessing ground truth parameters, our approach focuses solely on minimizing image error—a strategy intended to reduce parameter error indirectly. This method avoids the need for ground truth supervision, relying purely on images, which simplifies the learning process. Additionally, we do not meta-train initializations or step sizes, which could potentially offer further advantages depending on the specific inverse problem being addressed.

The meta-optimization over all soft rasterizers, parametrized by \softRendererParameters, is outlined as follows:
\smallskip
\[
\optimizer(
\mathbb E_i
[
||
\softRenderer(
\optimizer(
||
\softRenderer(
\sceneParameters, 
\softRendererParameters)
-
I_i||_2, 
\sceneParameters),
\softRendererParameters)
-
I_i||_2],
\softRendererParameters)
\,.
\]

These parametrized rasterizers differ from traditional renderers primarily in how they handle triangle edge testing and occlusions. Instead of using hard step functions, they employ a function, $\soft_\softRendererParameters(\distance)$, which is dependent on a specific parameter vector, \softRendererParameters. Next, we will explore various soft edge functions in further detail.

\begin{algorithm}[]
    \caption{\label{algo:algo}Meta-learning for Soft Rasterization}
    \begin{algorithmic}[1]
    \Require $\Theta$: Set of task images
    \Ensure Meta-learned soft renderer parameter \softRendererParameters
    \State \softRendererParameters = \Call{Random}{}
    \For{ $i \in [1, \stepRenderer] $}
    \State \sceneParameters = \Call{Random}{}
    \State $\sampleImage\leftarrow\Call{SampleImage}{\Theta}$ 
    \For{ $j \in [1, \stepScene] $}
    \State $
    \loss_\sceneParameters
    \leftarrow
    ||
    \softRenderer(\sceneParameters, \softRendererParameters)
    -
    \sampleImage
    ||_2
    $ 
    \State $\nabla_{\sceneParameters} = \Call{gradient}{\loss_\sceneParameters, \sceneParameters}$
    \State $\sceneParameters = \sceneParameters - \stepSizeScene\nabla_{\sceneParameters}$
    \EndFor 
    \State $\loss_\softRendererParameters \leftarrow
    ||
    \softRenderer(\sceneParameters, \softRendererParameters)
    -
    \sampleImage
    ||_2
    $
    \State $\nabla_{\softRendererParameters} = \Call{gradient}{}(\loss_\softRendererParameters,\softRendererParameters )$
    \State $\softRendererParameters = \softRendererParameters - \stepSizeRenderer\nabla_{\softRendererParameters}$ 
    \EndFor
    \end{algorithmic}
\end{algorithm}

\subsection{Tunable softness}

\begin{figure}[b]
    \centering
    \subfigure[3D shape]{
    \includegraphics[width=0.22\textwidth]{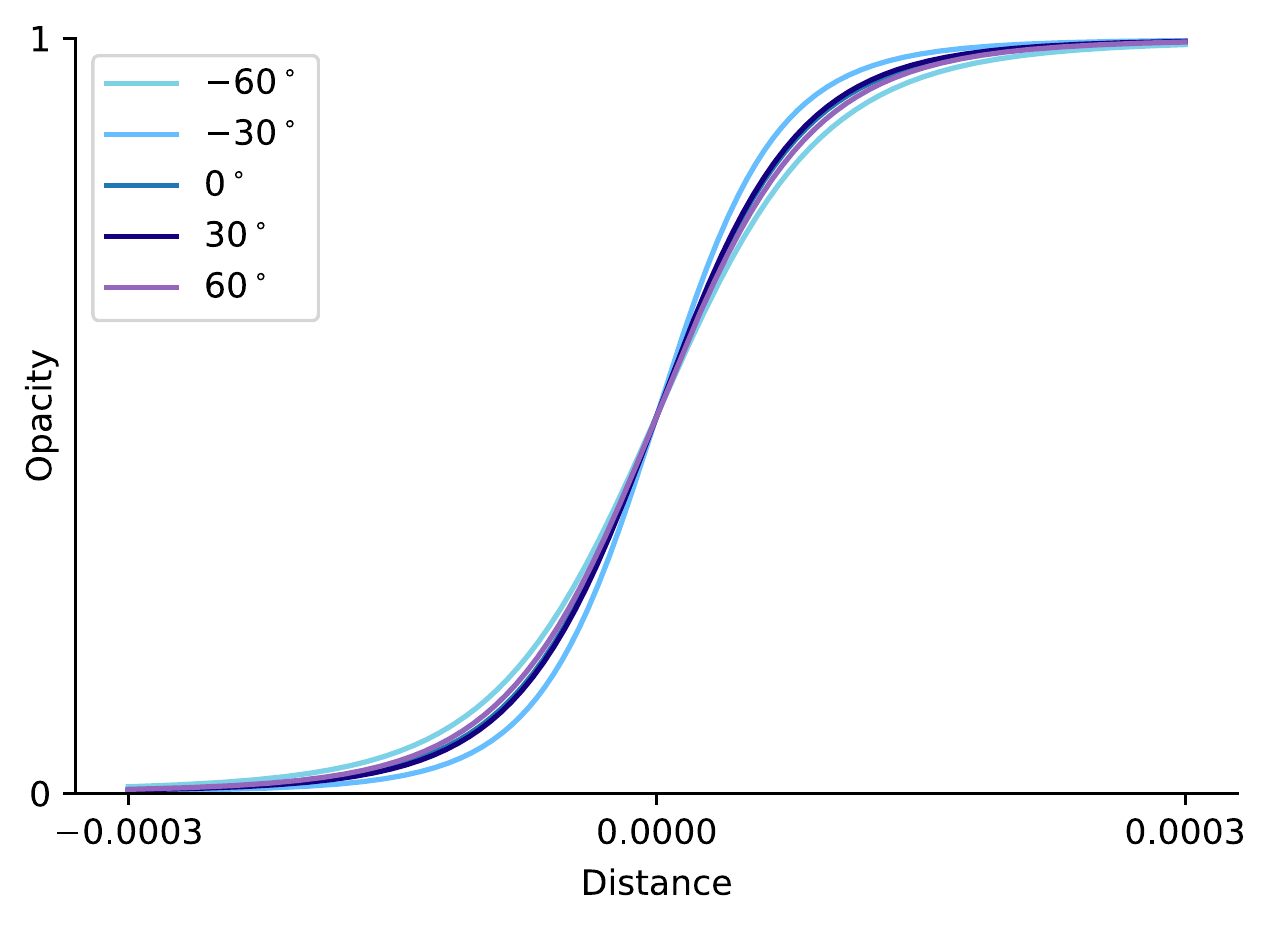}
}
    \subfigure[Pose]{
    \includegraphics[width=0.22\textwidth]{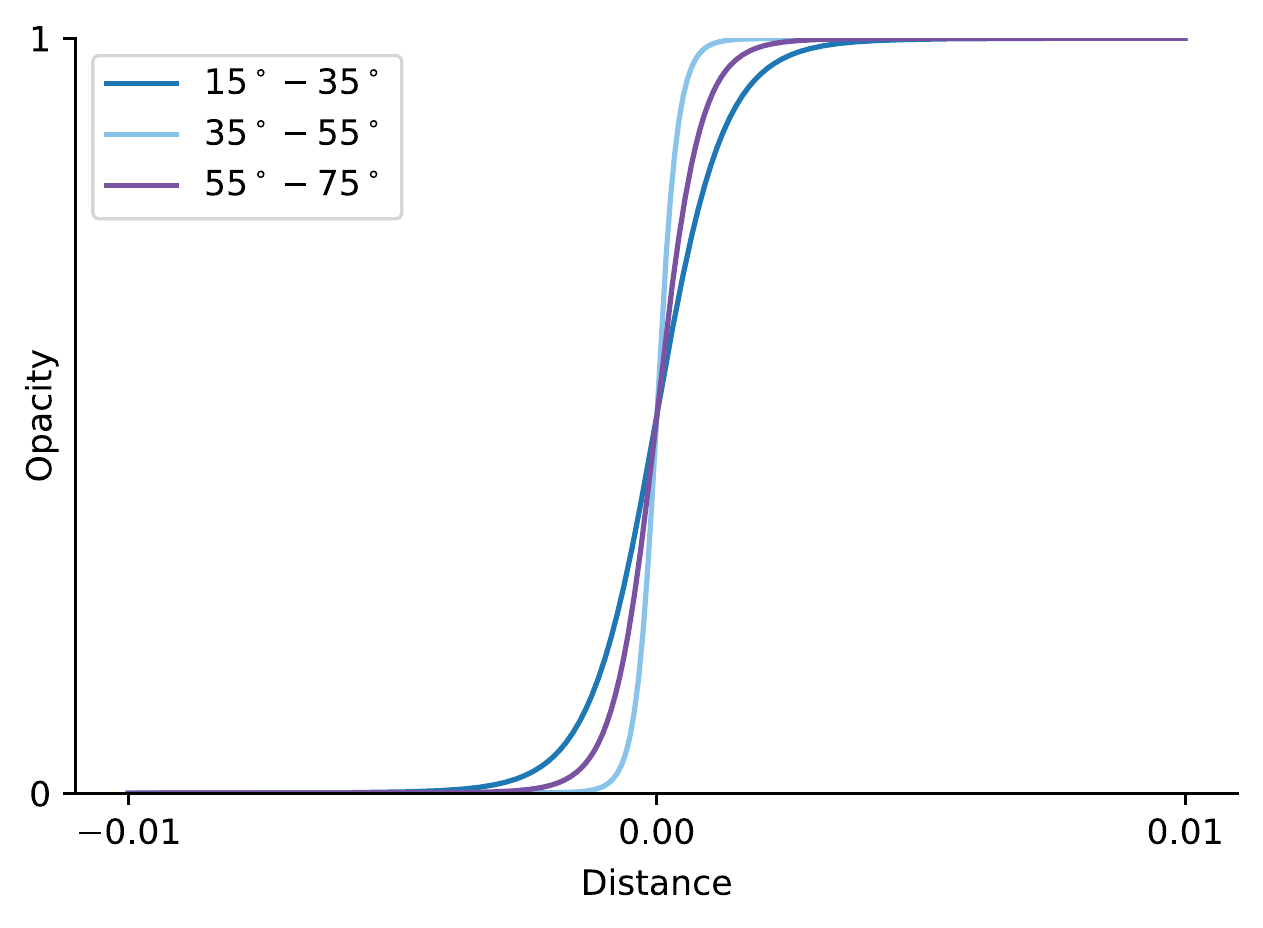}
}
    \subfigure[2D shape]{
    \includegraphics[width=0.22\textwidth]{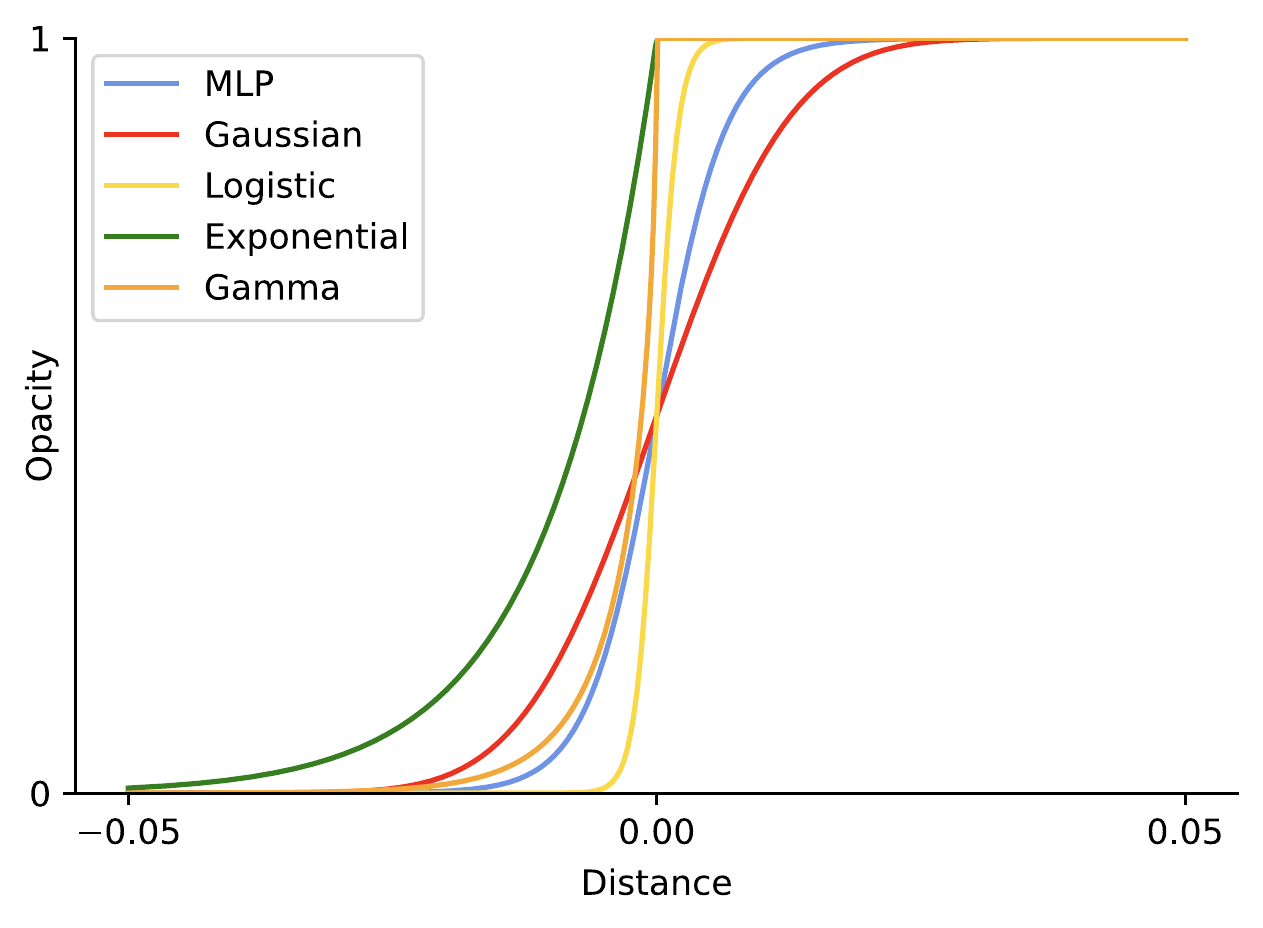}
}
    \subfigure[Occlusion]{
    \includegraphics[width=0.22\textwidth]{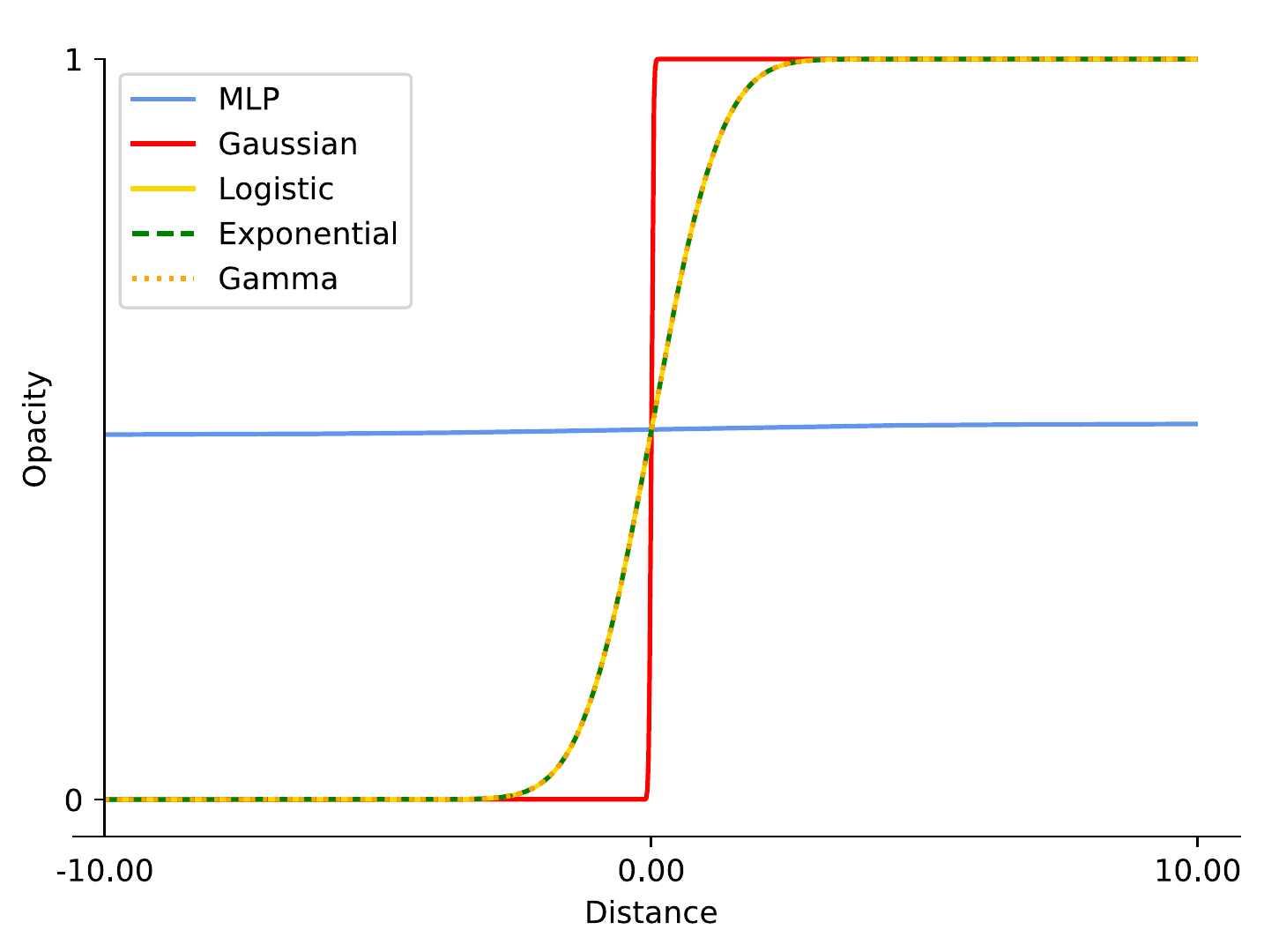}
}
    \caption{\label{fig:bluriness}
    \textbf{Edge functions for different tasks.}
    We visualize the meta-learned parameters in two sets of tasks. In 3D shape (a) and Pose (b), distinct MLPs are trained for each angle, shown in different shades of blue.
    For  2D shape (c) and Occlusion (d), the fixed viewpoint necessitates a single MLP, which can be compared to other CDFs with grid-searched softness. The same parameters can be used for more complex scenes (i.e. Transfer tasks) without re-meta-training.}
\end{figure}

We explore two types of tunable softness functions: \ac{CDF} and \acp{MLP}. Contrary to the approach taken by GenDR \cite{Petersen2022GenDR}, we do not strictly enforce the softening function to be an explicit \ac{CDF} of another function, as we find this requirement to be optional and not necessary for the function’s practical application.
Concurrently, with the flexibility of \ac{MLP}, more complex edge functions can be utilized that seem like variations of S-curves, but are actually optimized on the required task (Figure~\ref{fig:bluriness}). 

\mypara{CDF}
There are many options in this class of functions, such as a logistic function with a softening parameter \softRendererParameters
\smallskip
\begin{equation}
\soft_{\softRendererParameters_d}
=
\sigmoid(\distance)
=
\frac1{1 + \exp(-\distance\cdot\softRendererParameters)}
\,.
\end{equation}
\\
This class of functions has been previously utilized, allowing for parameter optimization through grid-search due to its low dimensionality. We further refine the optimization of \softRendererParameters using meta-learning, aiming for more precise control. Additionally, we find that a more general class of functions yields superior results, which we discuss next.

\mypara{MLP}
For more general softening, we employ an \ac{MLP} that comprises five layers with $\tanh$ as internal activations and a residual layer that skips three middle layers, followed by a final sigmoid:
\smallskip
\begin{equation}
\soft_{\softRendererParameters_d}(\distance)
=
\sigmoid(
\mathsf W_5
T(T(\mathsf W_1\distance)+
\mathsf W_4
T(\mathsf W_3
T(\mathsf W_2
T(\mathsf W_1 \distance)
))))
\,,
\end{equation}
where $T$ represents the $\tanh$ function. Note the repeated use of $\mathsf W_1$, which functions as a residual connection. The network weights, $\softRendererParameters=\{\mathsf W_1,\ldots,\mathsf W_5\}$, have a width of 4 and are randomly initialized from a uniform distribution. Our investigations show that the inclusion of biases or affine coordinates, such as $\mathsf{W}_1(\distance+1)$, does not contribute effectively, so we opt to exclude them from this configuration.

\subsection{Tunable depth}

To take the renderer's flexibility a step further, we consider the use of CDFs and MLP in the aggregation function, to evaluate our approach's performance in tasks like occlusion.
Originally, the sigmoid function is directly used on inverted normalized depth to calculate the contribution of each primitive's color in the final pixel. 
With the aim of making the depth itself differentiable, we first apply a soft function to the depth and then use it as the input of the sigmoid function. 
We use the CDFs as is, but some modifications are made to the structure of MLP. 
Since the sigmoid will be applied in the aggregation function (\cite{liu2019soft}), the last layer of the MLP can be removed. To ensure non-negative depth throughout differentiation as the camera is set to origin by default, the activation function is changed from $\tanh$ to $ReLU$. Keeping the same structure as our softness function, we have:
\smallskip
\begin{equation}
\soft_{\softRendererParameters_z}(\depth)
=
\mathsf W_5
R(R(\mathsf W_1\depth)+
\mathsf W_4
R(\mathsf W_3
R(\mathsf W_2
R(\mathsf W_1 \depth)
)))
\,,
\end{equation}
\smallskip
where $R$ represents the ReLU function.

\subsection{Combination}

Overall, our proposal involves meta-learning the corresponding parameters \softRendererParameters, namely $\softRendererParameters_\mathrm d$ for spatial distance and $\softRendererParameters_\mathrm z$ for depth, to enable soft blending in distance \distance and depth \depth.
We blend the colors of all primitives for a pixel with weights, that are made to sum to 1 by a normalization.
To this end, we define the final value at pixel $i$ as follows:
\smallskip
\begin{equation}
C_i=
\left(
\sum_j
\soft_{\softRendererParameters_d}(\distance_j^i)
\cdot
\soft_{\softRendererParameters_z}(\depth_j^i)
\right)^{-1}
\cdot
\sum_j
\soft_{\softRendererParameters_d}(\distance_j^i)
\cdot
\soft_{\softRendererParameters_z}(\depth_j^i)
\cdot
C^i_j
\end{equation}
\smallskip
where $C^i_j$ is the color of pixel $i$ at the $j$-th primitive.

\section{Evaluation}

We compare different methods (Sec.~\ref{sec:methods}) according to different metrics (Sec.~\ref{sec:metrics}) on different tasks (Sec.~\ref{sec:tasks}) leading to the results presented in Sec.~\ref{sec:results}. To keep consistency with our backbone, GenDR, we also use our MLP on the famous single-view 3D reconstruction experiment trained on the ShapeNet dataset (Sec.~\ref{sec:shapenet}). 

\begin{table}[t]
\centering
\small
\caption{Methods in our comparison experiments.} 
\label{tab:methods}
\renewcommand{\tabcolsep}{0.15cm}
\begin{tabular}{llcrl}
\toprule
\multicolumn1c{Methods}&
\multicolumn1c{Backbone}&
\multicolumn1c{Function}&
\multicolumn1c{Params}&
\multicolumn1c{Tuning}
\\
\midrule
\method{\textcolor{mlp}{MLP}}&
GenDR&
$\mathtt{MLP}(\distance;\softRendererParameters)$ &
56&
Meta-learned
\\
\method{\textcolor{gaussian}{Gauss}}&
GenDR&
$\mathcal{N}(\distance\cdot\softRendererParameters)$&
1&
Grid-searched
\\
\method{\textcolor{logistic}{Log}}&
GenDR&
$1/(1+\exp(-\distance\cdot\softRendererParameters))$&
1&
Grid-searched
\\
\method{\textcolor{exponential}{Exp}}&
GenDR&
$\exp(\distance;\softRendererParameters)$&
2&
Grid-searched
\\
\method{\textcolor{gamma}{Gamma}}&
GenDR&
$\mathcal{G}(\distance;\softRendererParameters;p$ = $0.5)$&
3&
Grid-searched
\\ 
\bottomrule
\end{tabular}
\end{table}

\subsection{Methods} \label{sec:methods}

We evaluate five different methods, as summarized in Tab.~\ref{tab:methods}. One method utilizes our meta-learned \ac{MLP}, while the others employ simpler edge functions with low-dimensional parameters suitable for grid-search optimization. For these existing edge functions, we use parameters that have been previously established in the literature through grid searches.
Except for the variations in soft edge handling, all methods employ identical rendering setups, including perspective projection and Phong materials, and operate at the same resolution without super-sampling. We use GenDR as the foundational backbone, into which we integrate various edge functions for comparison.

\subsection{Metrics} \label{sec:metrics}

While image distance serves as our primary loss, we analyze results using two distinct types of metrics as shown in the Metric of Tab.~\ref{tab:tasks}. The first is image distance: for the \task{2D Shape} and \task{Occlusion} tasks, this is measured using Mean Squared Error (MSE), while for \task{3D Shape} and \task{Pose}, we use Intersection over Union (IoU) \cite{Everingham2010}.

The second type of metric is parameter distance, which provides insights into the accuracy with which the models can determine underlying geometric or positional parameters. Specifically, for the \task{3D Shape} task, we use Chamfer distance; for \task{Pose}, we measure angle differences; and for \task{Occlusion}, we assess depth order. It's important to note that for \task{2D Shape}, identifying ground truth parameters for triangle vertices is challenging. Since our optimization efforts aim to uncover these parameters, this additional metric offers valuable insight into each method's performance.

Furthermore, it's worth noting that we did not utilize these parameters during the learning phase, and they're only used for comparison in evaluation. Our focus does not extend to optimizing light or materials, as they are out of the scope of rasterizers.
When presenting results, whether through charts or mean value, we ensure reliability by averaging data across 30 to 300 runs of the inner optimizer, depending on the task, in line with practices from GenDR. 

\begin{table}[b]
    \centering
    \caption{Properties of all tasks we study.}
    \setlength{\tabcolsep}{0.1cm}
    \label{tab:tasks}
    \begin{tabular}{ll rr cc ll}
    \toprule
        \multicolumn1c{Dim.}& 
        \multicolumn1c{Task}& 
        \multicolumn2c{DOF}& 
        \multicolumn1c{Dec.}& 
        \multicolumn1c{Occ.}&
        \multicolumn2c{Metric}\\
        \cmidrule(lr){3-4}
        \cmidrule(lr){7-8}
        &&
        \multicolumn1c{Original}&
        \multicolumn1c{Trans.}&
        &&
        \multicolumn1c{Im.}&
        \multicolumn1c{Para.}\\
        \midrule
         2D &
         \task{2D Shape} &
         800 &
         4,000 &
         \xmark &
         \xmark &
         MSE &
         \multicolumn1c{---}\\
         3D &
         \task{3D Shape}&
         3,840 & 
         3,840 &
         \xmark& 
         \xmark & 
         IoU & 
         Chamf.\\ 
         3D & 
         \task{Pose}& 
         4 &
         4 & 
         \cmark & 
         \xmark & 
         IoU & 
         Angle\\
         1D & 
         \task{Occlusion}& 
         3 & 
         8 & 
         \xmark & 
         \cmark& 
         MSE & 
         Depth\\ 
         \bottomrule
        \end{tabular}
\end{table}

\subsection{Tasks} \label{sec:tasks}

We explore four inverse rendering tasks for optimizing 2D and 3D shapes, camera pose, and occlusion as illustrated in Fig.~\ref{fig:bluriness}. For each task, we study two variants: the \emph{original} version and the \emph{transfer} one. 
In the \emph{original} setting, the rasterizer is meta-learned on the task with a designated input, say we find the best edge softness to make one particular logo. 
In the \emph{transfer} instance, the trained rasterizer is used on the same task but with different input and initial parameters, \eg a new logo with 4000 initial triangles instead of 800.
The initial parameters are pointed out in Tab.~\ref{tab:tasks}, under the ``Original'' and ``Trans.'' columns.
For some problems, we dynamically vary the size of spatial and depth distances. For other problems, we use preset values, following GenDR’s decay protocol where applicable (column ``Dec.'' in Tab.~\ref{tab:tasks}).
The ``Occ.'' column shows the use of soft functions for color aggregation and depth tuning, which is only utilized for \task{Occlusion} task.

\mypara{2D Shape}
In this task, we determine the positions of hundreds of 2D triangle vertices to recreate a specific target image. The initial positions of these triangles are randomized. For the \emph{original} variant, an EGSR logo serves as the target image during both the meta-learning and optimization phases. The challenge is to accurately reconstruct this logo. In the \emph{transfer} variant, we apply both meta-learned and grid-searched softness to a different EGSR logo, requiring optimization of additional initial triangles to reshape the target.

\mypara{3D Shape}
We begin by optimizing a mesh sphere to match the silhouette of an airplane within 100 steps. To evaluate this, we render the airplane from five different elevation angles $(([-60^\circ, -30^\circ, 0^\circ, 30^\circ, 60^\circ]))$ and sample 24 azimuthal views at each elevation, averaging the loss. We compare the performance of our MLP with meta-learned parameters against CDFs with grid-searched softness. While the \emph{original} variant uses an airplane, the \emph{transfer} variant tests the method on a chair.

\mypara{Pose}
In this scenario, the geometry remains fixed while we optimize the camera's pose; the up and look-at directions are set constants. The camera position is represented in spherical coordinates. We randomly initialize the viewing angle between $[10^{\circ},30^{\circ}]$ and distance from $[2,8]$. Azimuth and elevation are normally randomized from $[15^{\circ},75^{\circ}]$ and segmented into three parts for evaluation. For each segment, we calculate the average loss across 200 pairs of reference and initial images. An Adam optimizer ($\beta_1=0.9, \beta_2=0.99, \epsilon=10^{-8}$) with different learning rates (0.03 and 0.3) is employed for optimizing both the meta parameters and the camera pose. The \emph{original} variant involves a cow \cite{crane2013robust}, and the \emph{transfer} variant focuses on a dragon.

\mypara{Occlusion} 
Initially, we set three overlapping quadrilaterals with distinct colors and textures, and optimize their depths to achieve the correct occlusion order. Depths are uniformly randomized within $z \in [0.5,1.5]$. 
In the \emph{transfer} version, we move the focus of our task from ordering the quadrilaterals to correctly determining the closest face to the scene. The challenge in this version would be having more initial quadrilaterals with complex textures (eight overlapping ones with depths ranging from $z \in [-1.5,2]$), and finding the correct top face in fewer iterations. The MLP outperforms all the distributions by modifying the order correctly in just 10 iterations. 

This setup tests each method's ability to accurately process the anteroposterior relationship, especially when quadrilaterals fully overlap, challenging the capabilities of hard $z$-buffering and silhouette aggregation. Note that in our optimization process, only the $z$ coordinate of each primitive is adjusted, while the $x$ and $y$ coordinates are held fixed. 

\subsection{Implementation}

We implement the meta-learned soft rasterization with PyTorch, leveraging CUDA extensions to compute gradients precisely for high efficiency.
All the experiments run on a NVIDIA V100 SXM2 16GB GPU and we use a meta-learning technique that is similar to traditional \ac{MAML} to learn our meta parameters, which represent the softness of different distributions both in 2D and depth space.
The full algorithm is outlined in Algo.~\ref{algo:algo}.

Note that the optimization is not supervised by the ground truth parameters at any point.

In the \task{Shape} and \task{Pose} tasks, we apply a hard $z$-buffer.
In the \task{Occlusion} task, we use the same edge function for all variants of depth, to solely focus on their performance as soft depth functions.
In all tasks, we use probabilistic sum as silhouette aggregation \cite{liu2019soft} for all methods. 

\subsection{Results} \label{sec:results}

We report first quantitative and later qualitative results of our approach.

\begin{table*}[]
    \centering
    \caption{Image and parameter error for different tasks (columns) and different methods (rows).
    The best method is shown in \winner{bold} font.}
    \label{tab:results}
    \setlength{\tabcolsep}{0.16cm}
    \begin{tabular}{l rrrr rrrr rrrr rrrr}
        \toprule
        &
        \multicolumn4c{\task{2D Shape}}&
        \multicolumn4c{\task{3D Shape}}&
        \multicolumn4c{\task{Pose}}&
        \multicolumn4c{\task{Occlusion}}
        \\
        \cmidrule(lr){2-5}
        \cmidrule(lr){6-9}
        \cmidrule(lr){10-13}
        \cmidrule(lr){14-17}
        &
        \multicolumn2c{Original}&
        \multicolumn2c{Transfer}&
        \multicolumn2c{Original}&
        \multicolumn2c{Transfer}&
        \multicolumn2c{Original}&
        \multicolumn2c{Transfer}&
        \multicolumn2c{Original}&
        \multicolumn2c{Transfer}\\
        \cmidrule(lr){2-3}
        \cmidrule(lr){4-5}
        \cmidrule(lr){6-7}
        \cmidrule(lr){8-9}
        \cmidrule(lr){10-11}
        \cmidrule(lr){12-13}
        \cmidrule(lr){14-15}
        \cmidrule(lr){16-17}
        &
        \multicolumn1c{Im.}&
        \multicolumn1c{Para.}&
        \multicolumn1c{Im.}&
        \multicolumn1c{Para.}&
        \multicolumn1c{Im.}&
        \multicolumn1c{Para.}&
        \multicolumn1c{Im.}&
        \multicolumn1c{Para.}&
        \multicolumn1c{Im.}&
        \multicolumn1c{Para.}&
        \multicolumn1c{Im.}&
        \multicolumn1c{Para.}&
        \multicolumn1c{Im.}&
        \multicolumn1c{Para.}&
        \multicolumn1c{Im.}&
        \multicolumn1c{Para.}
        \\
        \midrule
\method{\textcolor{mlp}{MLP}} & \winner{0.0038} & \tabna & \winner{0.0038} & \tabna & \winner{19.4} & \winner{3.2} & \winner{69.0} & 4.72 & \winner{1.1} & \winner{0.01} & \winner{2.17} & 0.14 & \winner{0.041} & \winner{0.244} & \winner{0.040} &1.63 \\
\method{\textcolor{gaussian}{Gauss}} & 0.0046 & \tabna & 0.0187 & \tabna & 31.2 & 11.9 & 79.0 & 4.49 & 15.3 & 0.19 & 15.98 & 0.14 & 0.049 & 0.962 & 0.058 & 7.63 \\
\method{\textcolor{logistic}{Log}} & 0.0056 & \tabna & 0.0179 & \tabna & 29.8 & 10.1 & 97.0 & 4.54 & 16.2 & 0.22 & 16.63 & \winner{0.12} & 0.048 & 0.916 & 0.055 & 7.51 \\
\method{\textcolor{exponential}{Exp}} & 0.0054 & \tabna & 0.0148 & \tabna & 32.0 & 11.3 & 87.0 & \winner{4.35} & 15.9 & 0.18 & 14.98 & 0.15 & 0.049 & 0.627 & 0.046 & \winner{1.40} \\
\method{\textcolor{gamma}{Gamma}} & 0.0056 & \tabna & 0.0204 & \tabna & 28.7 & 10.0 & 86.0 & 4.58 & 14.5 & 0.24 & 15.14 & 0.24 & 0.049 & 0.627 & 0.046 & \winner{1.40} \\     
\bottomrule
\end{tabular}
\end{table*}

\begin{figure*}[t]
	\centering
	\addtolength{\tabcolsep}{-5pt}
	\hspace*{-10pt}
	\begin{tabular}{c cc cc}
        &
        \multicolumn{2}{c}{Original Tasks}
	    &
	    \multicolumn{2}{c}{Transfer Tasks}
	    \\
        &
        Image error &
	    Parameter error &
	    Image error &
	    Parameter error
        \\

        \raisebox{20pt}{\rotatebox{90}{\task{2Dshape}}}
        &
	    \includegraphics[width=0.235\textwidth]{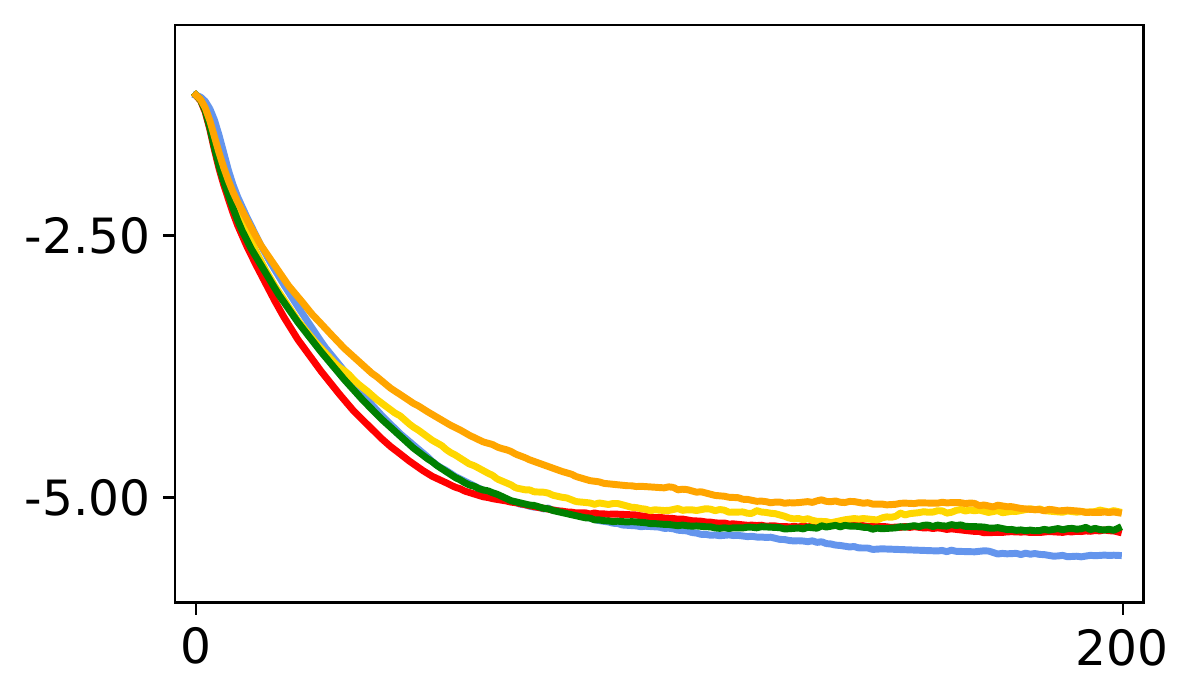} 
        &

        \raisebox{1.2cm}{N/A}
        &
	    \includegraphics[width=0.235\textwidth]{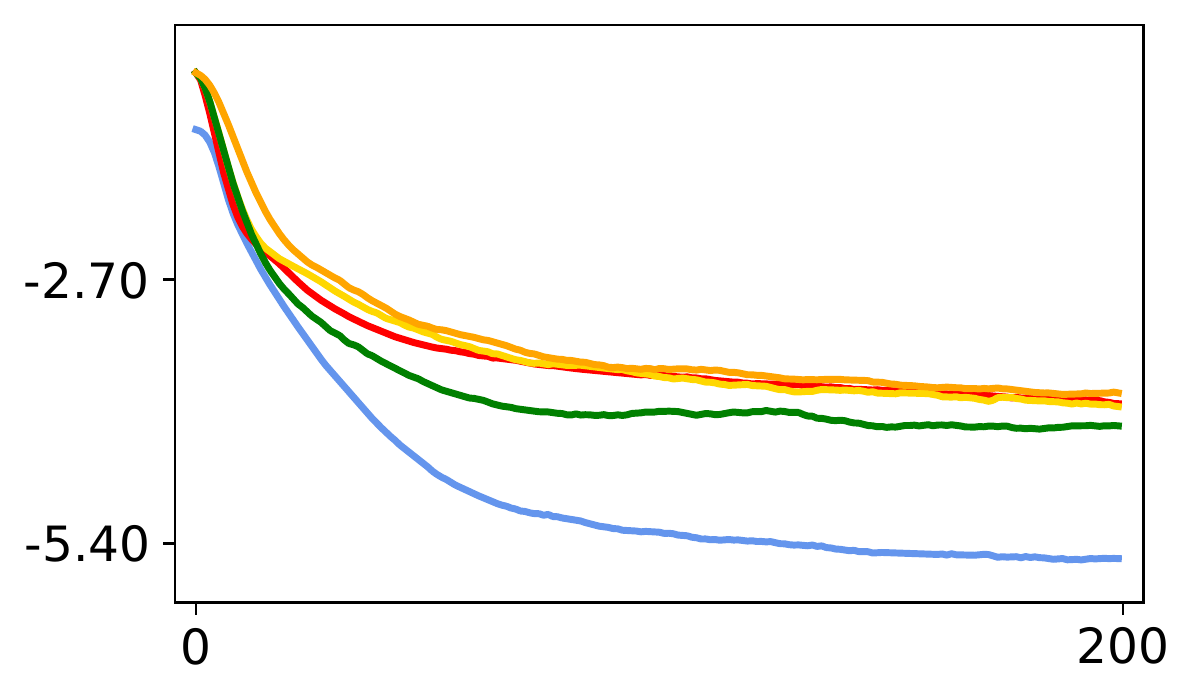} 
        &

        \raisebox{1.2cm}{N/A}
        \\

        \raisebox{15pt}{\rotatebox{90}{\task{3Dshape}}}
        &
	    \includegraphics[width=0.235\textwidth]{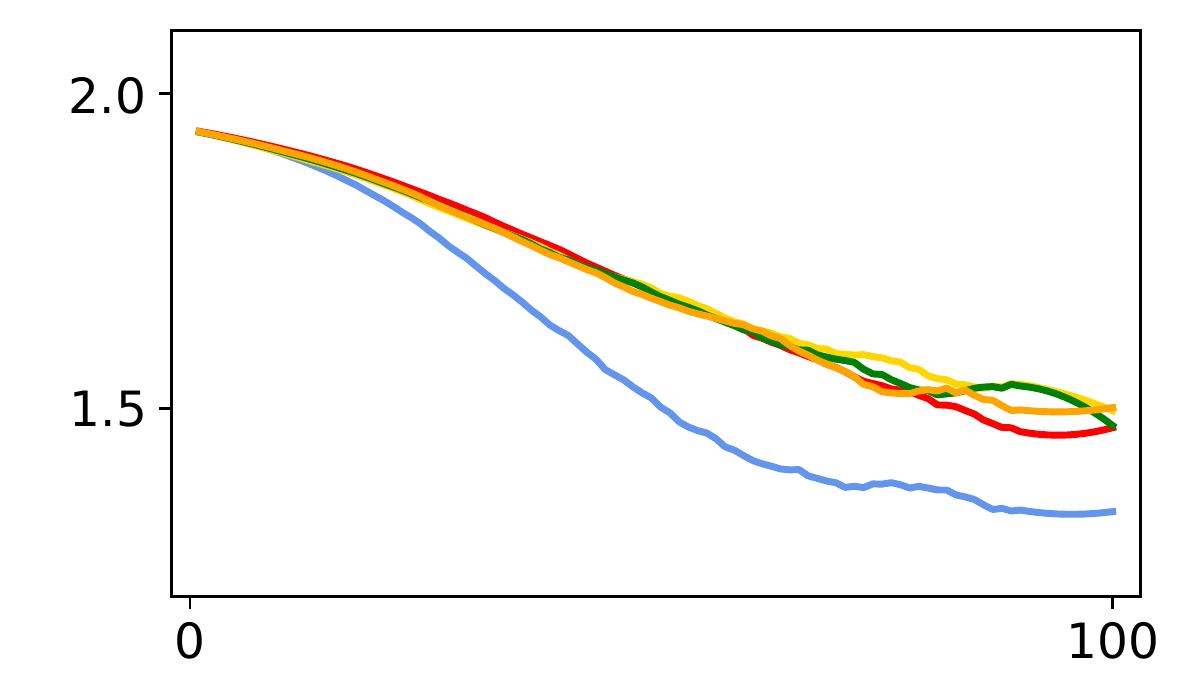} &
	    \includegraphics[width=0.235\textwidth]{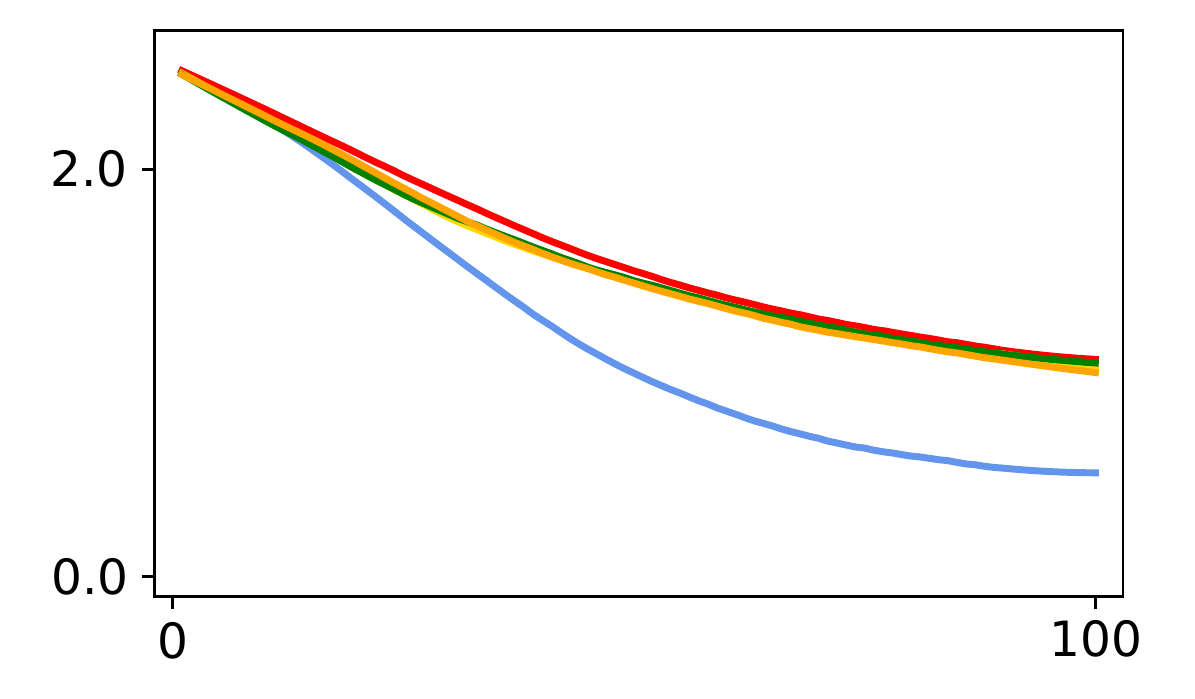} &
	    \hspace*{0.2cm}\includegraphics[width=0.235\textwidth]{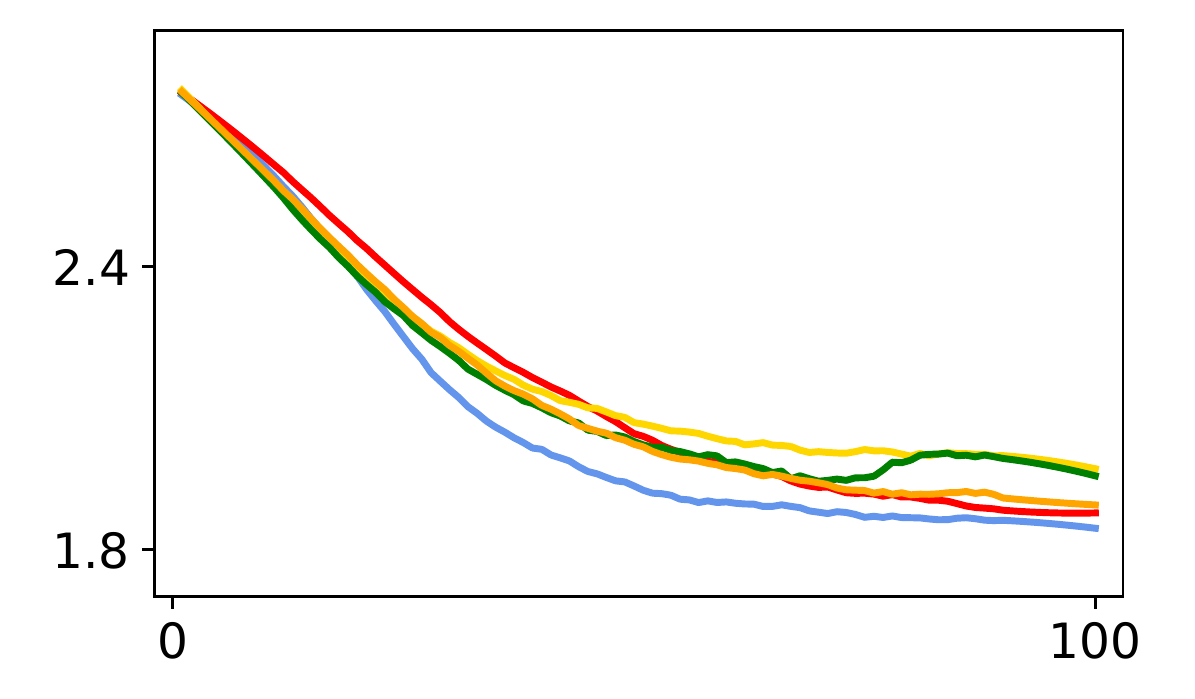} &
	    \hspace*{0.1cm}\includegraphics[width=0.235\textwidth]{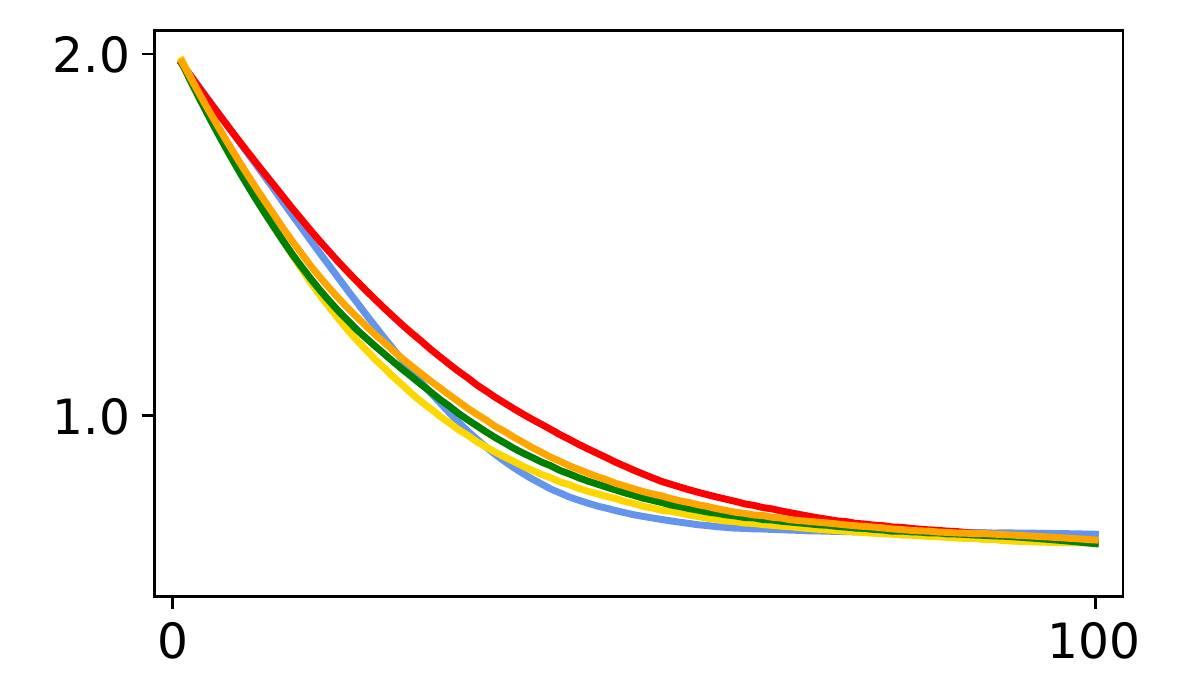}  
        \\

        \raisebox{25pt}{\rotatebox{90}{\task{Pose}}}
        &
	    \includegraphics[width=0.235\textwidth]{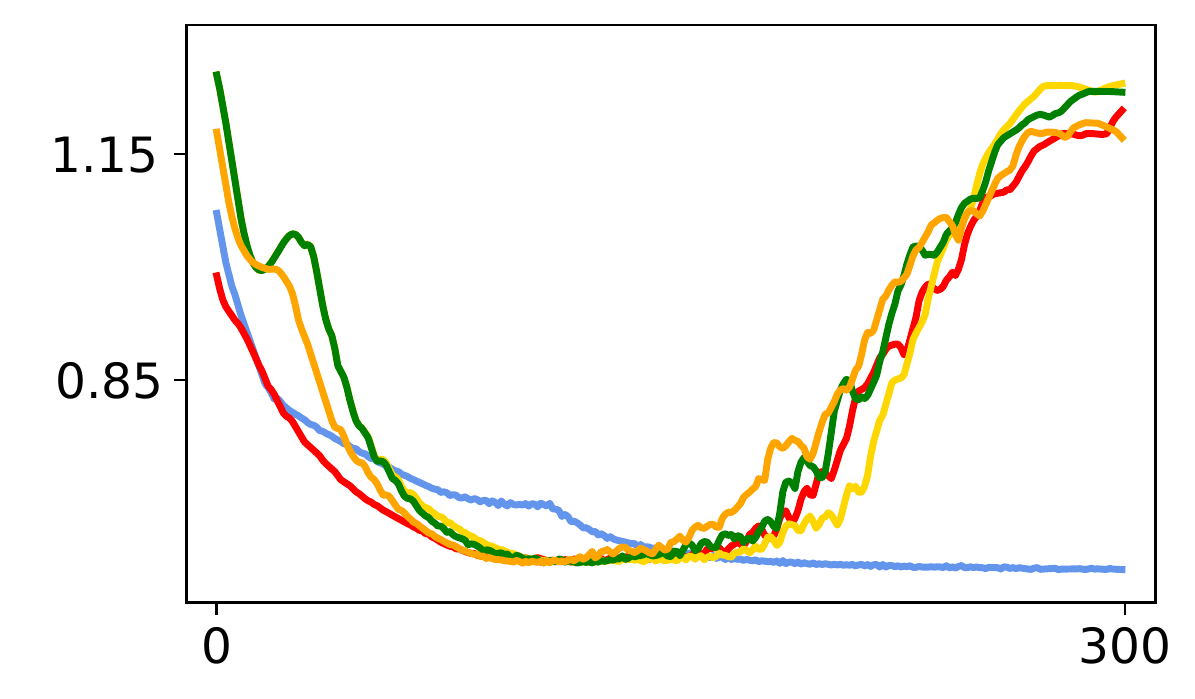} &
	    \includegraphics[width=0.235\textwidth]{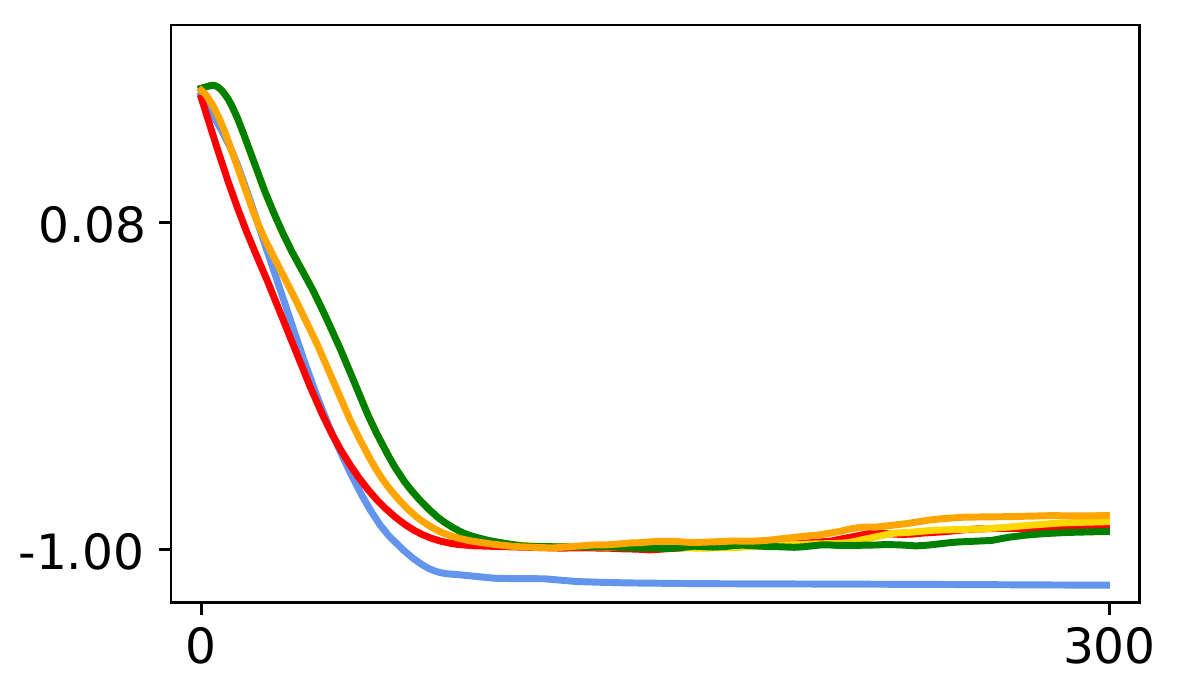} &
	    \includegraphics[width=0.235\textwidth]{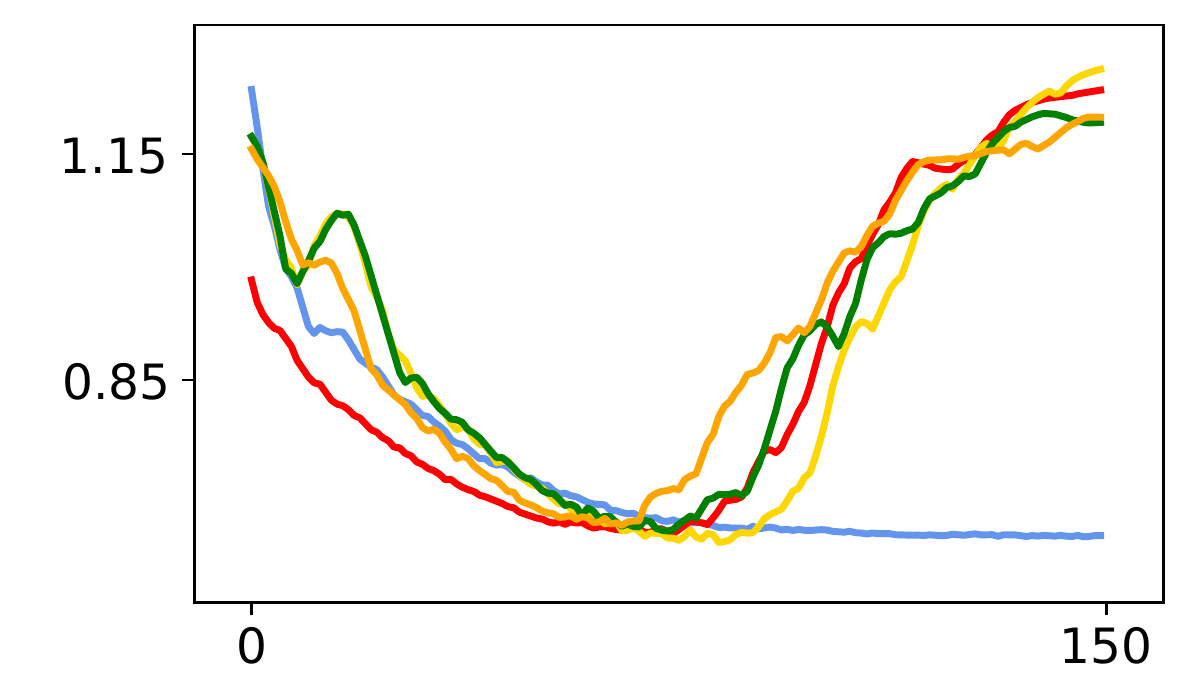} &
	    \includegraphics[width=0.235\textwidth]{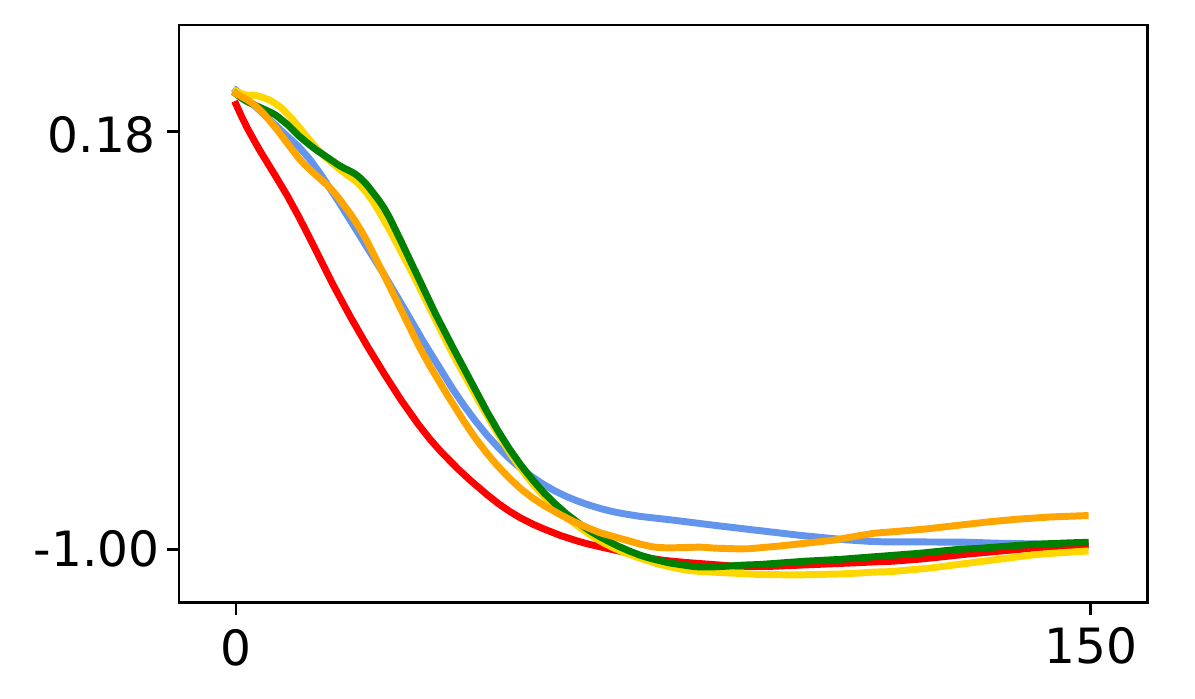}  
        \\
        \raisebox{15pt}{\rotatebox{90}{\task{Occlusion}}}
        &
        \includegraphics[width=0.235\textwidth]{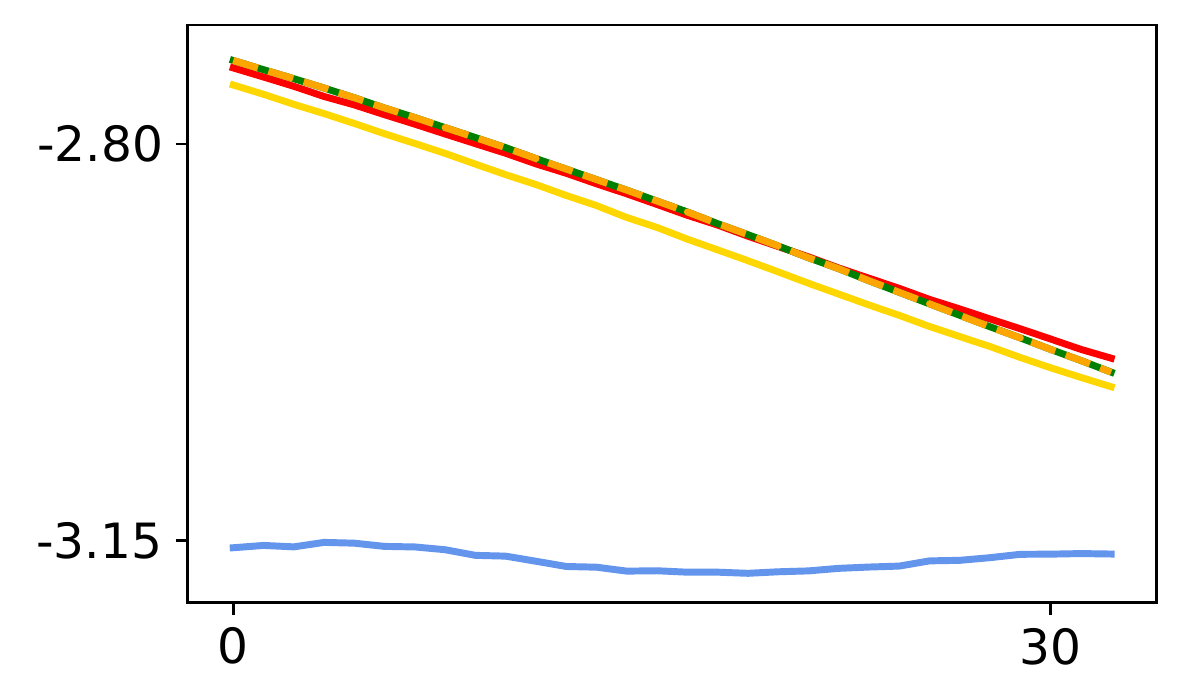} &
	    \includegraphics[width=0.235\textwidth]{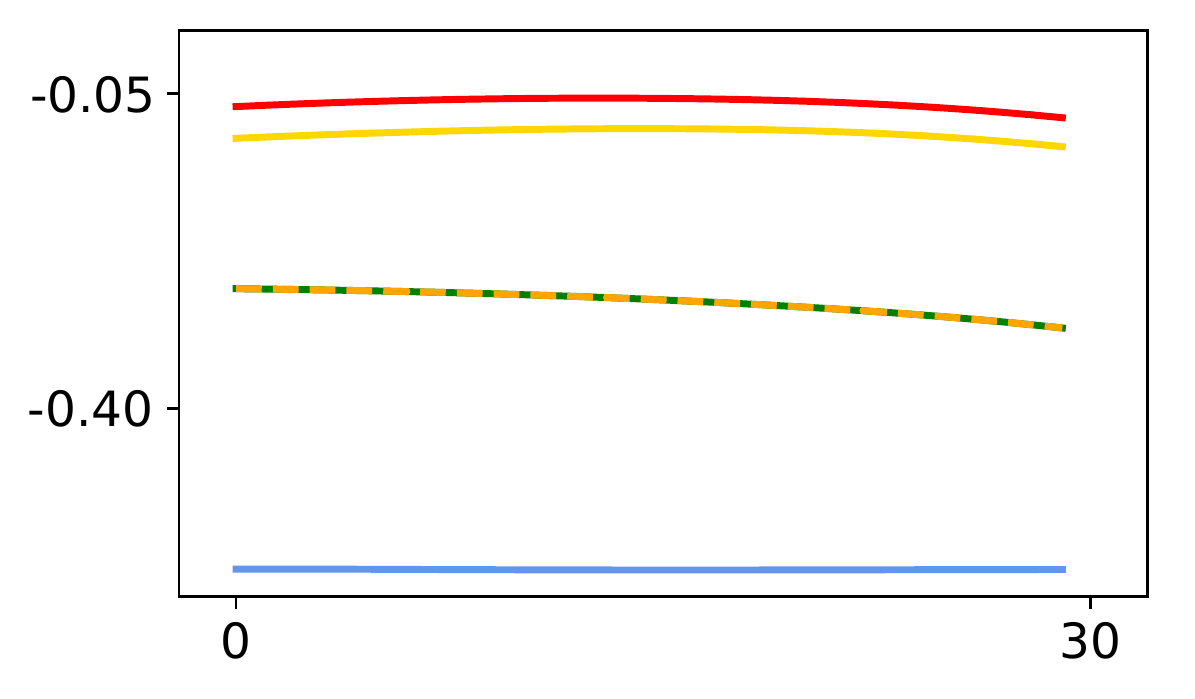} &
	    \includegraphics[width=0.235\textwidth]{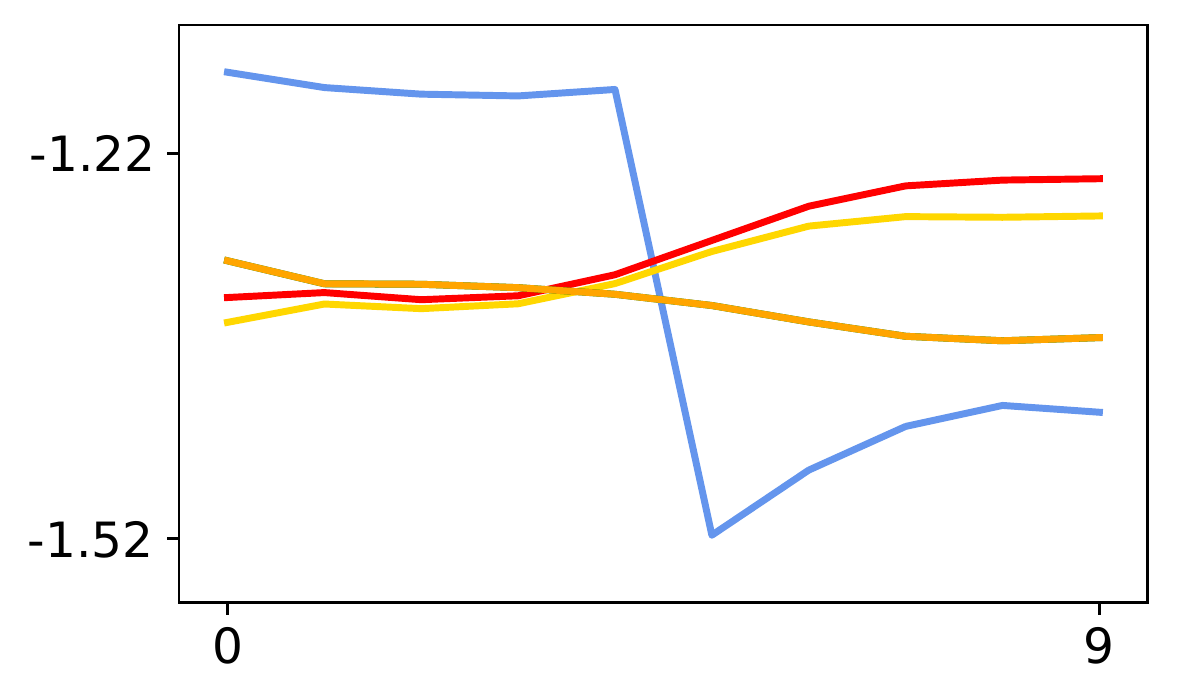} &
	    \includegraphics[width=0.235\textwidth]{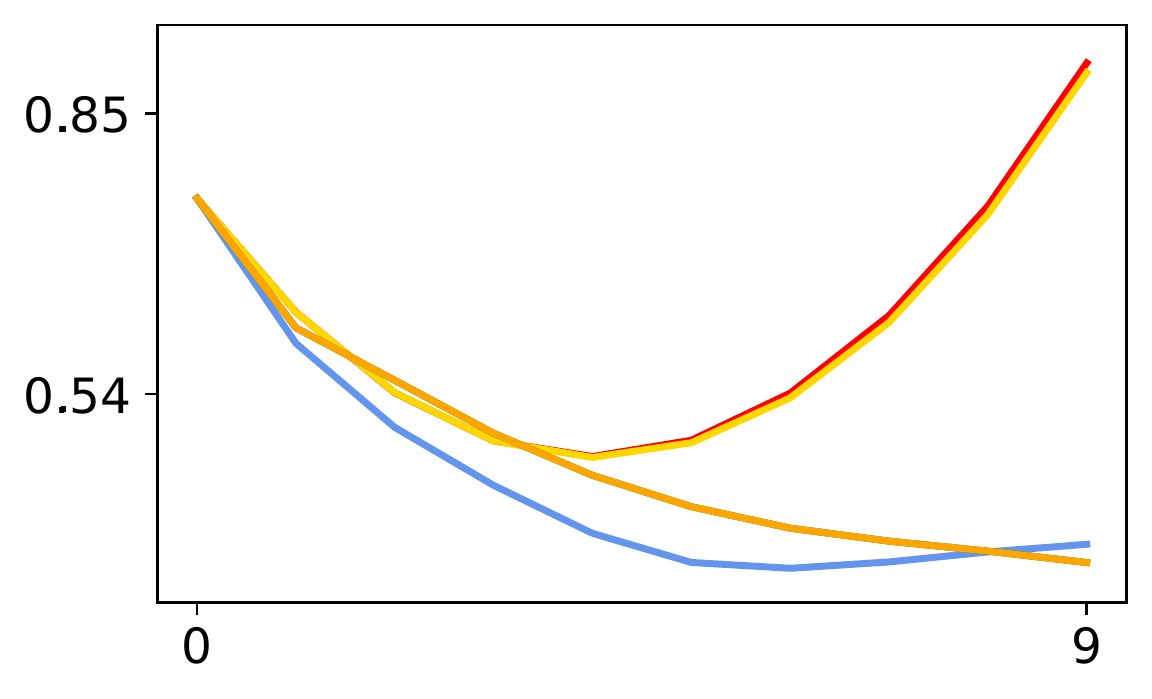}  
	    \\
        \multicolumn{5}{c}{\includegraphics[width=0.5\textwidth]{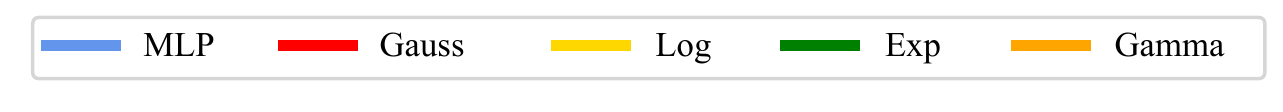}}

	\end{tabular}
    \caption{
    Every subplot shows the \textbf{convergence of one inverse rendering task} according to one metric where different colors represent different methods.
    Within each subplot the vertical axis is loss, so less is better (log scale).
    The horizontal axis is optimization iterations.
    The first two columns show a training variant, the last two columns show a transfer condition.
    In each horizontal pair, the first plot is the image-based metric, the second one is the parameter error.
    }
    \label{fig:loss_plots}
\end{figure*}

\mypara{Quantitative}
Quantitative results are seen in Fig.~\ref{fig:loss_plots} and Tab.~\ref{tab:results}, where we analyze all tasks according to all metrics using all methods.
In Fig.~\ref{fig:loss_plots}, a successful method will ---according to both metrics--- have a graph that quickly and reliably goes to a low error value; and also stays down.
In Tab.~\ref{tab:results}, lower is better.

We see that across the tasks, and consistent between metrics, our \method{MLP} performs best (blue).
This is true both for the endpoint (right in each plot and values in Tab.~\ref{tab:results}), as for most (interruptible) in-between iterations as well.
For some iterations (horizontal axis in each plot), all methods perform similarly in most tasks and for both metrics, but eventually, meta-learned methods take the lead while others plateau.
In some examples, existing methods could not solve the task with the published default values at all, while ours can adapt to any task on any scale.
We also see, that while our work optimizes the image error, the unseen parameter also converges to its lowest, which is the ultimate objective in reverse rendering tasks.

For \task{Occlusion}, MLP performs consistently better compared to other distributions. While a convergence seems possible in the original version based on the image, the parameter error shows otherwise, meaning that the image error is not enough for these distributions to learn depth.

In all experiments, the transfer from learning on one task and testing on another can succeed, as deduced from comparing the first and the second pair of rows, in which the first task class is seen in meta-training, while the second one of the pair is not. 
This shows the potential to save computational resources on optimizing complex tasks by training on related simpler ones. 

\begin{table*}
    \centering
    \caption{\textbf{Comparison of 3D IoU for \task{single-view 3D mesh reconstruction} in GenDR and our method.} The results from GenDR are the best results reported in their paper, comparing a total of 30 renderers, while ours represent a single instance of renderer with MLP as the distribution function.}
    \setlength{\tabcolsep}{0.16cm}
    \label{tab:shapenet}
    \begin{adjustbox}{width=\textwidth}
    \begin{tabular}{lllllllllllllll}
    \toprule
    Method               & Airplane & Bench  & Dresser         & Car    & Chair  & Display & Lamp   & Speaker         & Rifle  & Sofa   & Table  & Phone  & Vessel & Mean   \\ \hline
    GenDR\cite{Petersen2022GenDR} (best results) & 0.6473   & 0.5026 & 0.7175          & 0.7696 & 0.5297 & 0.6147  & 0.4665 & 0.6673          & 0.6773 & 0.6879 & 0.4961 & 0.8189 & 0.6006 & 0.6232 \\
    MLP + Probabilistic  & 0.6385   & 0.4755 & \textbf{0.7216} & 0.6849 & 0.5086 & 0.6033  & 0.4594 & 0.6671 & 0.6613 & 0.6449 & 0.4617 & 0.7890 & 0.5829 & 0.6076 \\ 
    Difference           & 0.0088     & 0.0271   & -0.0041            & 0.0847   & 0.0211   & 0.0114    & 0.0071   & 0.0002            & 0.0160   & 0.0430   & 0.0344   & 0.0299   & 0.0177   & 0.0156  \\
    \bottomrule
    \end{tabular}
    \end{adjustbox}
\end{table*}

\begin{figure*}
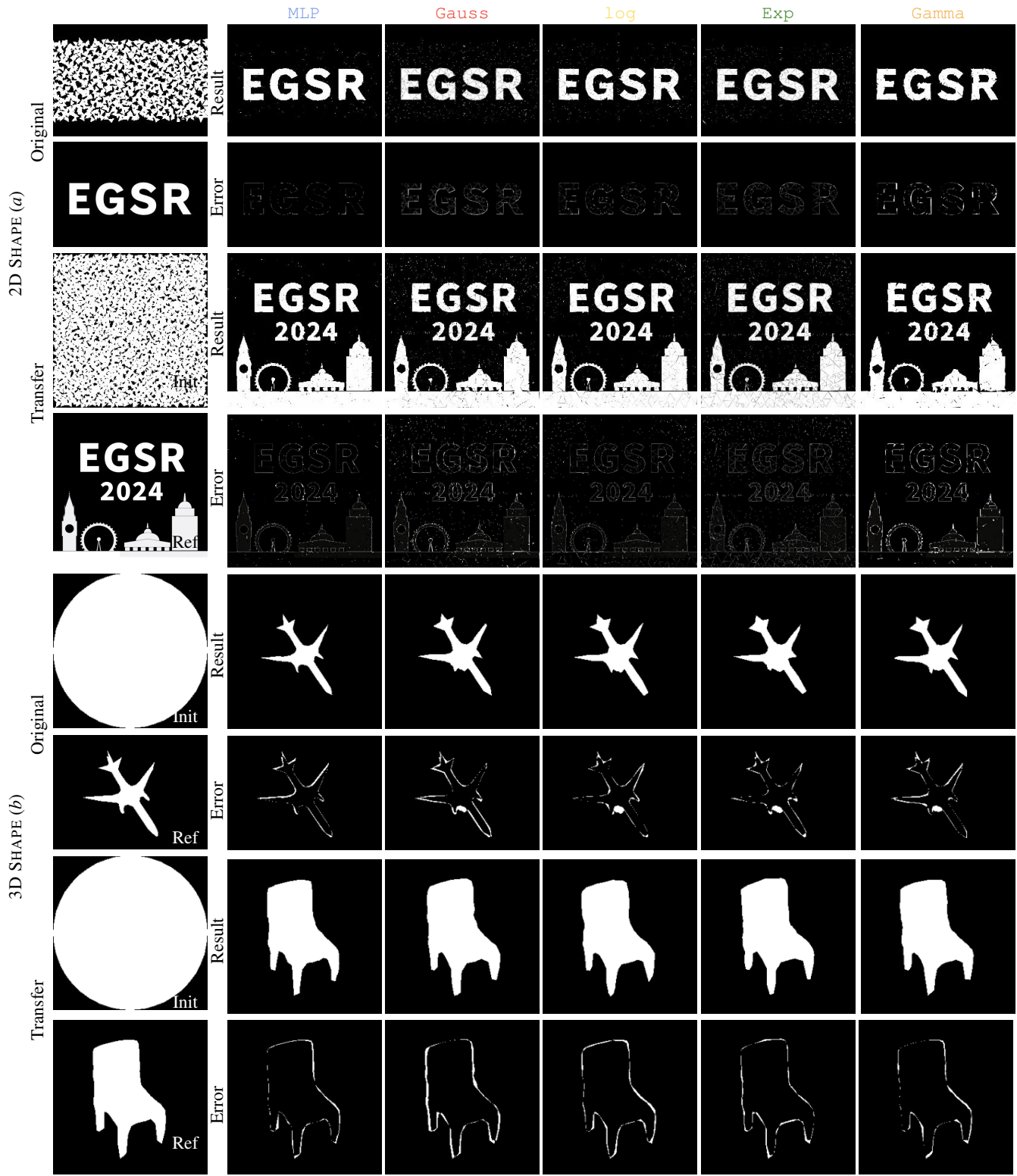

	\centering

	\addtolength{\tabcolsep}{-5pt}

	\begin{tabular}{ccccccccc}
	&&&&
    \method{\textcolor{mlp}{MLP}} & 
    \method{\textcolor{gaussian}{Gauss}} & 
	\method{\textcolor{logistic}{log}} &
	\method{\textcolor{exponential}{Exp}} &
	\method{\textcolor{gamma}{Gamma}}
	\\
    \input{images/final_results/shape2D/shape2D}
    \\
    \input{images/final_results/shape3D/shape3D}

	\end{tabular}
	
    \caption{
    \label{fig:shape2d_shape3d_results}
    \textbf{Results of different methods for the \task{2D Shape} and \task{3D Shape} task.}
    Every pair of rows is one task. The first two pairs are 2D tasks, the second two pairs are 3D tasks.
    Every even rows show a rendering of the result upon convergence, except for the first column, where we show the initialization (random tris in 2D tasks and a sphere in 3D tasks).
    Every odd row, shows the error image of that, except for the first column, which shows the reference, the target.
    A successful optimization would have a black error image and a result that looks similar to the reference. Please refer to the supplementary materials for more analysis.
    }
\end{figure*}

\begin{figure*}
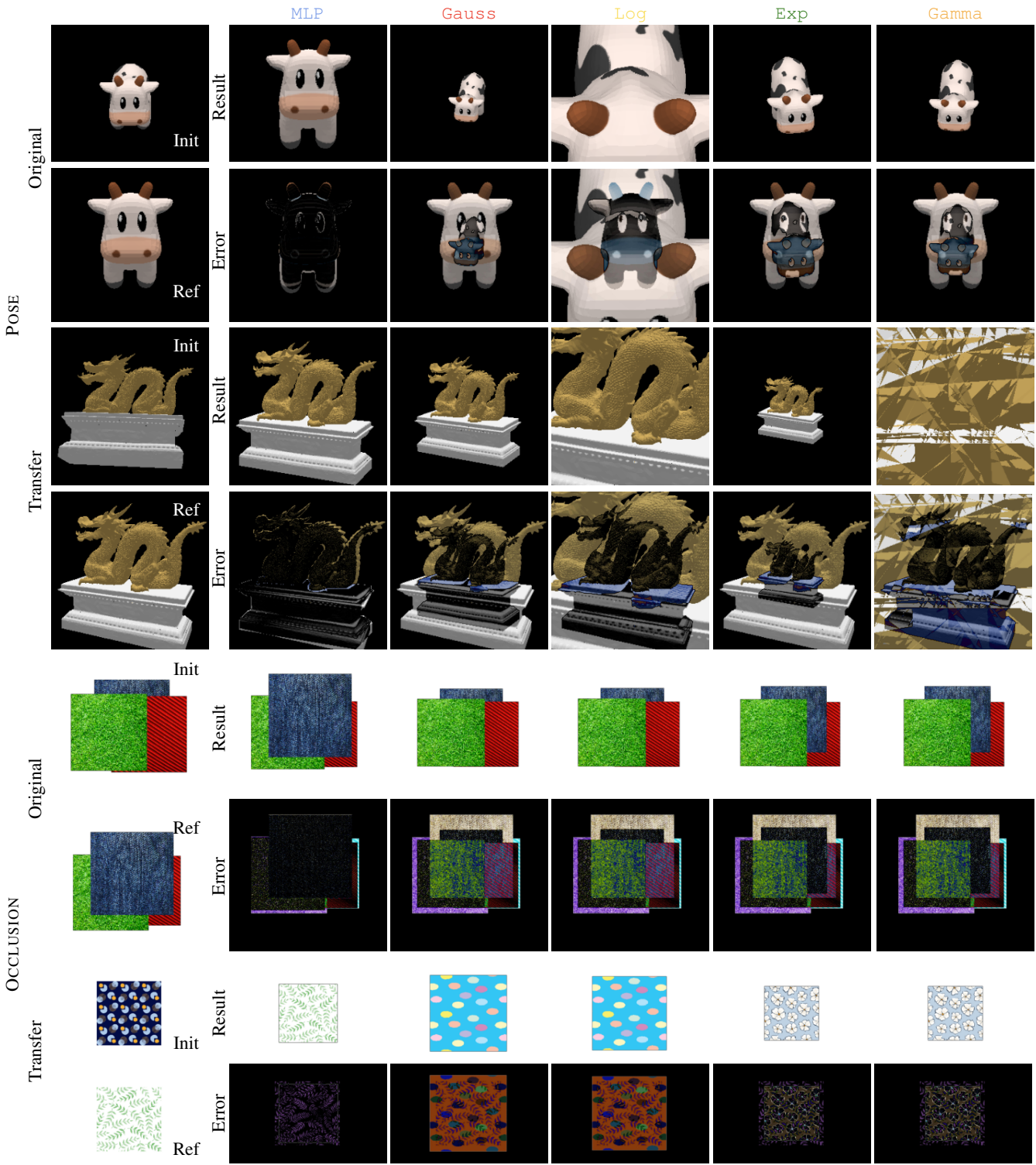

	\centering

	\addtolength{\tabcolsep}{-5pt}

	\begin{tabular}{ccccccccc}
	&&&& 
    \method{\textcolor{mlp}{MLP}} & 
    \method{\textcolor{gaussian}{Gauss}} & 
	\method{\textcolor{logistic}{Log}} &
	\method{\textcolor{exponential}{Exp}} &
	\method{\textcolor{gamma}{Gamma}}
	\\
    \input{images/final_results/pose/pose}
    \\
    \input{images/final_results/occlusion/occlusion}

	\end{tabular}
	
    \caption{
    \label{fig:pose_occlusion_results}
    \textbf{Results of different method for the \task{Pose} and \task{Occlusion} task.}
    Every pair of rows is one task. The first two pairs are \task{Pose} tasks, the second two pairs are occlusion tasks.
    Every even rows show a rendering of the result upon convergence, except for the first column, where we show the initialization (random poses in the \task{Pose} task and random quads in the \task{Occlusion} task).
    Every odd row, shows the error image of that, except for the first column, which shows the reference, the target.
    A successful optimization would have a black error image and a result that looks similar to the reference. Please refer to the supplementary materials for more analysis.}
\end{figure*}

\mypara{Qualitative}
Similarly, the qualitative results of the same tasks are illustrated in Fig.~\ref{fig:shape2d_shape3d_results}, and Fig.~\ref{fig:pose_occlusion_results}, comparing the renderings of the final iteration from Fig.~\ref{fig:loss_plots}.
As shown in the Error results, \method{MLP} performs best.
For \task{Shape} and \task{Pose} tasks, our results match the reference better than the previous methods starting from the same initialization.  Additionally, the successful transfer of meta-parameters from one task to another is evident in each pair of columns. Despite the substantial differences in 3D shapes and discrepancies in DOF, the meta-learned softness effectively facilitates a transfer and achieves the lowest error.

Note that for \task{Pose} task, GenDR gives unsatisfactory results, which are faithfully obtained by directly running the code released by them. 
The reason why their optimization ended up so bad is that they manually set the range of dynamic softness from $10^{-1}$ to $10^{-7}$.
As shown in the supplementary materials, their method achieves optima when the softness decays to about $10^{-5}$, but it finally gets worse as the softness keeps decaying and becomes improper.

This is one of the drawbacks of manually designed softness, even if GenDR sets higher softness at the early stage and decreases it as the solution converges, it’s still unknown when and where the softness reaches the optima. However, our meta-learned softness does not have such problems and limitations but also shows a coarse-to-fine transition in practice (if set to higher initially).

\subsection{Single-View 3D Mesh Reconstruction} \label{sec:shapenet}

For this final experiment, we follow the same auto-encoder structure, as proposed in all preceding studies (\cite{Kato2018mesh,liu2019soft, chen2019_dibr, petersen2021learning, Petersen2022GenDR}), and constrict the number of renderers to one with the distribution function using MLP and probabilistic T-conorm. 
Since the implementation of nested loops is not time-efficient due to the task's complexity, we optimize the parameters of our MLP with the same Adam optimizer used for the model's parameters. While we have the disadvantage of not reaching convergence for our MLP, in Tab.~\ref{tab:shapenet}, we show that our results are comparable with the best results reported in GenDR, which is searched over a large set of distribution functions and T-Conorms.

\subsection{Time Performance}

For the time performance of training, taking the \task{3D Shape} task as an example, GenDR employs coarse-to-fine grid-searching in 28 iterations. Our method meta-learns an MLP in 50 iterations but achieves better softness and results without manual range or fineness presumptions.

For the time performance of evaluation, our method only replaces a CDF with a small MLP with 56 parameters, which only contributes to less than 1$\%$ of the whole computational graph. 

These claims can be further proved by comparing the computational costs of training, evaluating, and rendering. The time costs (seconds) corresponding to Tab.~\ref{tab:results} are shown in Tab.~\ref{tab:timecost}.

From Tab.~\ref{tab:timecost} we can see that the MLP’s extra time cost is negligible compared to the whole Evaluate Time, which includes forward rendering, backward pass, loss calculation, etc.
As analyzed above, the time cost of MLP compared to GenDR for other tasks is similar. For Sec.~\ref{sec:shapenet}, the extra cost introduced by MLP is less and our method will be faster than grid-searching across 30 different renderers - while being slightly slower than every single renderer.
\begin{table}[h]
\centering
\small
\caption{The time(s) cost of each method in \task{3D shape}.} 
\label{tab:timecost}
\begin{tabular}{crrrr}
\toprule
Distribution & Train & Evaluate & Render & Image Loss \\
\midrule
\method{\textcolor{mlp}{MLP}} & 229.2 & 4.36 & 0.088 & 69.0 \\
\method{\textcolor{gaussian}{Gauss}} & 74.95 & 3.24 & 0.053 & 79.0 \\
\method{\textcolor{logistic}{Log}} & 83.9 & 3.34 & 0.059 & 97.0 \\ 
\method{\textcolor{exponential}{Exp}} & 87.7 & 3.51 & 0.057 & 87.0 \\
\method{\textcolor{gamma}{Gamma}} & 94.5 & 3.63 & 0.056 & 86.0 \\
\bottomrule
\end{tabular}
\end{table}

\subsection{Cross Evaluation}
\begin{table}[b]
\centering
\small
\caption{The loss of using softness meta-trained from each task for other tasks.} 
\label{tab:crosstask}
\begin{tabular}{ c rrrr}
\toprule
&\multicolumn4c{Train}\\
\cmidrule(lr){2-5}
Evaluate& \task{2D Shape} & \task{3D Shape} & \task{Occlusion} & \task{Pose} \\ 
\midrule
\task{2D Shape} & \winner{1 $\times$ 14.7} & 1.3 $\times$ 19.4 & 9.9 $\times$ 0.042 & 1.8 $\times$ 2.4\\ 
\task{3D Shape} & 1.2 $\times$ 14.7 & \winner{1 $\times$ 19.4} & 9.9 $\times$ 0.042& 1.4 $\times$ 2.4\\ 
\task{Occlusion} & 53.7 $\times$ 14.7 & 2.1 $\times$ 19.4 & \winner{1 $\times$ 0.042} & 1.7 $\times$ 2.4 \\
\task{Pose} & 1.1 $\times$ 14.7 & 1.1 $\times$ 19.4 & 9.9 $\times$ 0.042 & \winner{1 $\times$ 2.4} \\
\bottomrule
\end{tabular}

\end{table}
Tab.~\ref{tab:crosstask} shows what happens when a softness meta-trained from one task is used for another task. The rows stand for the tasks on which MLPs are trained, the columns refer to the evaluated tasks.

The lowest image loss for each task is on the diagonal, which means a softness learned on one task works better on this task than any other task, and this is true for all tasks, respectively, showing that meta-learning is effective. Note that here we use static softness for \task{Pose} task.

\section{Conclusion and discussion}

We introduced the application of meta-learning to acquire knowledge from a continuous space of softening operations. This approach is used to soften edges and occlusion functions, which in turn improves differentiable rendering. Our approach utilizes a tunable \ac{MLP} for space and depth edge functions, allowing for joint optimization of their parameters alongside scene parameters. This addresses issues related to discontinuities, enables the optimization of shape and pose, and resolves occlusions in depth.

Additionally, we have investigated the generalization capabilities of meta-learned softening operations, demonstrating the potential of our method to tackle complex rendering problems. The results underscore the transformative impact of our approach on complex differentiable rendering techniques.

Furthermore, we have explored adapting softening functions dynamically based on the task, moving away from a one-size-fits-all approach to one that adjusts softness based on specific scene requirements. Compared to manually designing the range and decreasing step size, our method shows more robustness and adaptability. Both static and dynamic strategies of softness can work, but it's a trade-off. Dynamic softness could be more precise but also needs more computations.

In future work, we plan to extend this technique to broader applications where discontinuities in integrands hinder differentiation and optimization, such as in physical differential equations and dynamic optimization. We envision meta-learning different softness for different optimization stages (from earlier to later) to capture different frequency detail levels.
This strategy suggests that softening could evolve into a more high-dimensional and complex process as the dynamic transition has not yet been proven to be necessarily monotonic, potentially making it better suited for implementation with an \ac{MLP}.
Similarly, the combination strategy of multiple softness in multi-task optimization is also a good direction to explore.

Moreover, this technique can be further improved by using the neural representation or neural proxy methods to provide heuristic gradients for the rasterization instead of analytical approaches as explored in this work.

\bibliographystyle{eg-alpha-doi} 
\bibliography{egbibsample}       

\newcommand{\etalchar}[1]{$^{#1}$}
\begin{thebibliography}{\uppercase{EVGW{\etalchar{*}}10}}

\bibitem[BLD20]{bangaru2020warpedsampling}
\textsc{Bangaru S., Li T.-M., Durand F.}:
\newblock Unbiased warped-area sampling for differentiable rendering.
\newblock \emph{ACM Trans. Graph. 39}, 6 (2020), 245:1--245:18.

\bibitem[CLG{\etalchar{*}}19]{chen2019_dibr}
\textsc{Chen W., Ling H., Gao J., Smith E., Lehtinen J., Jacobson A., Fidler S.}:
\newblock Learning to predict 3d objects with an interpolation-based differentiable renderer.
\newblock In \emph{NeurIPS} (2019).

\bibitem[CLG{\etalchar{*}}21]{Wenzheng2021DIB}
\textsc{Chen W., Litalien J., Gao J., Wang Z., Tsang C.~F., Khamis S., Litany O., Fidler S.}:
\newblock {DIB-R++}: Learning to predict lighting and material with a hybrid differentiable renderer, 2021.

\bibitem[CPS13]{crane2013robust}
\textsc{Crane K., Pinkall U., Schr{\"o}der P.}:
\newblock Robust fairing via conformal curvature flow.
\newblock \emph{ACM Transactions on Graphics (TOG) 32}, 4 (2013), 1--10.

\bibitem[EVGW{\etalchar{*}}10]{Everingham2010}
\textsc{Everingham M., Van~Gool L., Williams C. K.~I., Winn J., Zisserman A.}:
\newblock The pascal visual object classes (voc) challenge.
\newblock \emph{International Journal of Computer Vision 88}, 2 (2010), 303--338.

\bibitem[FAL17]{finn2017MAML}
\textsc{Finn C., Abbeel P., Levine S.}:
\newblock Model-agnostic meta-learning for fast adaptation of deep networks.
\newblock In \emph{ICLR} (2017), pp.~1126--1135.

\bibitem[FR23]{fischer2022plateau}
\textsc{Fischer M., Ritschel T.}:
\newblock Plateau-reduced differentiable path tracing.
\newblock In \emph{CVPR} (2023).

\bibitem[GMCG20]{Gupta2020pose}
\textsc{Gupta A., Medhi J., Chattopadhyay A., Gupta V.}:
\newblock End-to-end differentiable {6DoF} object pose estimation with local and global constraints, 2020.

\bibitem[KBM{\etalchar{*}}20]{Kato2020survey}
\textsc{Kato H., Beker D., Morariu M., Ando T., Matsuoka T., Kehl W., Gaidon A.}:
\newblock Differentiable rendering: A survey, 2020.

\bibitem[KUH18]{Kato2018mesh}
\textsc{Kato H., Ushiku Y., Harada T.}:
\newblock Neural {3D} mesh renderer.
\newblock In \emph{CVPR} (2018), pp.~3907--3916.

\bibitem[LADL18]{Li2018MC}
\textsc{Li T.-M., Aittala M., Durand F., Lehtinen J.}:
\newblock Differentiable monte carlo ray tracing through edge sampling.
\newblock \emph{ACM Trans. Graph. (Proc. SIGGRAPH) 37}, 6 (2018).

\bibitem[LB14]{Loper2014OpenDR}
\textsc{Loper M.~M., Black M.~J.}:
\newblock Opendr: An approximate differentiable renderer.
\newblock In \emph{ECCV} (2014), pp.~154--169.

\bibitem[LHJ19]{Loubet2019Reparameterizing}
\textsc{Loubet G., Holzschuch N., Jakob W.}:
\newblock Reparameterizing discontinuous integrands for differentiable rendering.
\newblock \emph{Transactions on Graphics (Proceedings of SIGGRAPH Asia) 38}, 6 (Dec. 2019).
\newblock \href {https://doi.org/10.1145/3355089.3356510} {\path{doi:10.1145/3355089.3356510}}.

\bibitem[LHK{\etalchar{*}}20]{nvidia2020}
\textsc{Laine S., Hellsten J., Karras T., Seol Y., Lehtinen J., Aila T.}:
\newblock Modular primitives for high-performance differentiable rendering.
\newblock \emph{ACM Trans. Graph. 39}, 6 (nov 2020).
\newblock URL: \url{https://doi.org/10.1145/3414685.3417861}, \href {https://doi.org/10.1145/3414685.3417861} {\path{doi:10.1145/3414685.3417861}}.

\bibitem[LLCL19]{liu2019soft}
\textsc{Liu S., Li T., Chen W., Li H.}:
\newblock Soft rasterizer: A differentiable renderer for image-based 3d reasoning.
\newblock In \emph{CVPR} (2019), pp.~7708--7717.

\bibitem[NDDJK21]{Nimier-David2021material}
\textsc{Nimier-David M., Dong Z., Jakob W., Kaplanyan A.}:
\newblock Material and lighting reconstruction for complex indoor scenes with texture-space differentiable rendering.
\newblock In \emph{Proc. EGSR} (2021), pp.~73--84.

\bibitem[PBDC19]{Petersen2019Pix2Vex}
\textsc{Petersen F., Bermano A.~H., Deussen O., Cohen{-}Or D.}:
\newblock Pix2vex: Image-to-geometry reconstruction using a smooth differentiable renderer.
\newblock \emph{CoRR abs/1903.11149} (2019).
\newblock \href {http://arxiv.org/abs/1903.11149} {\path{arXiv:1903.11149}}.

\bibitem[PBKD21]{petersen2021learning}
\textsc{Petersen F., Borgelt C., Kuehne H., Deussen O.}:
\newblock Learning with algorithmic supervision via continuous relaxations.
\newblock In \emph{NeurIPS} (2021).

\bibitem[PGBD22]{Petersen2022GenDR}
\textsc{Petersen F., Goldluecke B., Borgelt C., Deussen O.}:
\newblock {GenDR}: A generalized differentiable renderer.
\newblock In \emph{CVPR} (June 2022), pp.~4002--4011.

\bibitem[Pin88]{Pineda1988APA}
\textsc{Pineda J.}:
\newblock A parallel algorithm for polygon rasterization.
\newblock \emph{SIGGRAPH} (1988).

\bibitem[RRR{\etalchar{*}}15]{Rhodin2015AVS}
\textsc{Rhodin H., Robertini N., Richardt C., Seidel H.-P., Theobalt C.}:
\newblock A versatile scene model with differentiable visibility applied to generative pose estimation.
\newblock \emph{ICCV} (2015), 765--773.

\bibitem[YBAF22]{Yang2022AAF}
\textsc{Yang Y., Barnes C., Adams A., Finkelstein A.}:
\newblock A$\delta$: Autodiff for discontinuous programs – applied to shaders.
\newblock In \emph{ACM SIGGRAPH, to appear} (Aug. 2022).

\bibitem[ZMY{\etalchar{*}}20]{Zhang2020pathspace}
\textsc{Zhang C., Miller B., Yan K., Gkioulekas I., Zhao S.}:
\newblock Path-space differentiable rendering.
\newblock \emph{ACM Trans. Graph. 39}, 4 (2020).

\end{thebibliography}



\end{document}